\documentclass{IEEEtran}
\usepackage{cite}
\usepackage{amsmath,amssymb,amsfonts}
\usepackage{algorithmic}
\usepackage{graphicx}
\usepackage{textcomp}
\usepackage{multirow}
\usepackage{booktabs}
\usepackage{subfig}

\def\BibTeX{{\rm B\kern-.05em{\sc i\kern-.025em b}\kern-.08em
    T\kern-.1667em\lower.7ex\hbox{E}\kern-.125emX}}
\begin{document}

\title{A Survey on Nonconvex Regularization Based Sparse and Low-Rank Recovery in Signal Processing, Statistics, and Machine Learning}
\author{Fei Wen, \IEEEmembership{Member, IEEE}, Lei Chu, Peilin Liu, \IEEEmembership{Senior Member, IEEE}, and Robert Qiu, \IEEEmembership{Fellow,~IEEE}
\thanks{Department of Electronic Engineering, Shanghai Jiao Tong University, Shanghai, China (e-mail:
wenfei@sjtu.edu.cn; leochu@sjtu.edu.cn; liupeilin@sjtu.edu.cn; rcqiu@sjtu.edu.cn.)}}


\markboth{\underline{Published in IEEE Access: \MakeLowercase{https://ieeexplore.ieee.org/abstract/document/8531588}}}%
{}


\maketitle

\begin{abstract}
In the past decade, sparse and low-rank recovery have drawn much attention in many areas such as signal/image processing, statistics, bioinformatics and machine learning. To achieve sparsity and/or low-rankness inducing, the $\ell_1$ norm and nuclear norm are of the most popular regularization penalties due to their convexity. While the $\ell_1$ and nuclear norm are convenient as the related convex optimization problems are usually tractable, it has been shown in many applications that a nonconvex penalty can yield significantly better performance. In recent, nonconvex regularization based sparse and low-rank recovery is of considerable interest and it in fact is a main driver of the recent progress in nonconvex and nonsmooth optimization. This paper gives an overview of this topic in various fields in signal processing, statistics and machine learning, including compressive sensing (CS), sparse regression and variable selection, sparse signals separation, sparse principal component analysis (PCA), large covariance and inverse covariance matrices estimation, matrix completion, and robust PCA. We present recent developments of nonconvex regularization based sparse and low-rank recovery in these fields, addressing the issues of penalty selection, applications and the convergence of nonconvex algorithms. Code is available at https://github.com/FWen/ncreg.git.
\end{abstract}


\section{Introduction}
\label{sec:introduction}
{I}{n} the past decade, sparse and low-rank recovery have attracted much study attention in many areas,
such as signal processing, image processing, statistics, bioinformatics and machine learning.
To achieve sparsity and low-rankness promotion, sparsity and low-rankness constraints or penalties are commonly employed.
Among the sparsity and low-rankness inducing penalties, the $\ell_1$-norm
and nuclear norm penalties are of the most popular. This is mainly due to their convexity as
it makes the involved optimization problems tractable in that, existing convex optimization techniques
with well-established convergence properties can be directly used or can be applied after some extension.

Generally, under certain conditions, the $\ell_1$ and nuclear norm penalties can reliably recover
the underlying true sparse signal and low-rank matrix with high probability. However,
both of them have a bias problem, which would result in significantly biased estimates,
and cannot achieve reliable recovery with the least observations [1], [2], [3].
In comparison, a nonconvex penalty, such as the $\ell_0$, $\ell_q$ ($0<q<1$),
smoothly clipped absolute deviation (SCAD) or minimax concave penalty (MCP),
is superior in that it can ameliorate the bias problem of the $\ell_1$-one.
In recent, nonconvex regularization based sparse and low-rank recovery have drawn
considerable interest and achieved significant performance improvement in many applications
over convex regularization. This progress benefited a lot from the recent developments
in nonconvex and nonsmooth optimization, and, meanwhile, promoted the developments of the latter.

The goal of this article is to give an overview of the recent developments in
nonconvex regularization based sparse and low-rank recovery in signal/image processing,
statistics and machine learning. In a field as wide as this,
we mainly focus on the following eight important topics.

\hangafter 1 \hangindent 1.em
\textit{1) Compressive sensing (CS)}:
CS aims to acquire sparse signals (or signals can be sparsely represented in some basis)
at a significantly lower rate than the classical Nyquist sampling [14]--[16].
In CS, exploiting the sparsity (or sparse representation) of the desired signals
is the key for their reconstruction.
There exist a number of recent works addressing nonconvex regularized sparse reconstruction,
e.g., using the ${\ell _0}$ and ${\ell _q}$ regularization [39]--[66].
It has been demonstrated that, nonconvex regularization not only ameliorates the bias problem
of the ${\ell _1}$-one but also requires fewer measurements for reliable recovery.

\hangafter 1 \hangindent 1.em
\textit{2) Sparse regression and variable selection}:
Sparse regression aims to simultaneously select variables and estimate coefficients of variables,
which is a fundamental problem in high-dimensional statistical analysis. Nonconvex penalties,
such as the SCAD, ${\ell _0}$, ${\ell _q}$, MCP penalties,
have been widely employed to attain more accurate estimation over the ${\ell_1}$-one [11], [12], [83]--[96].

\hangafter 1 \hangindent 1.em
\textit{3) Sparse signals separation and image inpainting}:
Sparse signals separation problems arise in many important applications,
such as source separation, super-resolution, inpainting, interference cancellation,
saturation and clipping restoration, and robust sparse recovery in impulsive (sparse) noise.
In these applications, the ${\ell _q}$ penalty can attain considerable restoration
performance improvement over the ${\ell _1}$ penalty [120].

\hangafter 1 \hangindent 1.em
\textit{4) Sparse principal component analysis (PCA)}:
PCA is a useful statistical tool for data analysis and dimensionality reducing,
which is widely used in virtually all areas of science and engineering.
Sparse PCA aims to obtain sparse loading vectors to enhance the
interpretability of the principle components (PCs).
Nonconvex regularization, such as ${\ell _0}$ regularization,
has been widely used for sparsity promotion [128]--[141].

\hangafter 1 \hangindent 1.em
\textit{5) Large sparse covariance matrix estimation}:
Large covariance matrix estimation is a fundamental problem in
modern high-dimensional statistical analysis,
which has found wide applications such as in economics, finance,
bioinformatics, social networks, climate studies, and health sciences.
In the high-dimensional setting where the dimensionality is often
comparable to (or even larger than) the sample size,
sparsity regularized estimation is especially effective.
Some works addressing nonconvex regularized covariance matrix estimation include [148], [152].

\hangafter 1 \hangindent 1.em
\textit{6) Large sparse inverse covariance matrix estimation}:
Large inverse covariance matrix estimation is also fundamental to
high-dimensional statistical analysis, which is closely related to
undirected graphs under a Gaussian model.
Some papers addressing nonconvex regularized inverse covariance matrix estimation include [160]--[166].

\hangafter 1 \hangindent 1.em
\textit{7) Matrix completion}:
Matrix completion deals with the recovery of a low-rank matrix from
its partially observed entries. Such problems arise in many applications
such as in recommender systems, computer vision and system identification.
Various algorithms have been developed using the nonconvex Schatten-$q$ ($0 \le q < 1$) norm,
truncated nuclear norm, and MCP penalties for low-rankness inducing,
e.g., [6], [48], [175]--[182], to achieve better recovery performance over the nuclear norm.

\hangafter 1 \hangindent 1.em
\textit{8) Robust PCA}:
Robust PCA aims to enhance the robustness of PCA against outliers or
sparse corruption which is ubiquitous in modern applications such as
image processing, web data analysis, and bioinformatics.
Basically, robust PCA is a joint sparse and low-rank recovery problem,
which seeks to identify a low-dimensional structure from grossly corrupted observations.
Existing works using different combinations of nonconvex sparse and
low-rank penalties for robust PCA include [9], [193]--[197].

Among these topics, CS, sparse regression, sparse signals separation,
and sparse PCA are sparse vector recovery problems,
large sparse covariance and inverse covariance matrices estimation
are sparse matrix recovery problems,
whilst matrix completion and robust PCA are low-rank recovery problems.
To be more precise, sparse PCA is not a vector recovery problem,
but in many popular greedy methods, the PCs are estimated in a vector-by-vector manner.
Meanwhile, robust PCA is a joint sparse and low-rank recovery problem.

In this paper, we also provide some critical perspectives.
As it is often the case that, when new techniques are introduced,
there may also be some overexcitement and abuse.
We comment on the following points:
There exist certain instances in both sparse and low-rank recovery,
where the use of nonconvex regularization is simply unnecessary and
will not significantly improve performance [95].
The use of nonconvex regularization does not always guarantee distinct
performance improvement over convex regularization.
Moreover, employing nonconvex regularization models can even be
disadvantageous because the related nonconvex and nonsmooth
optimization problems are less tractable than convex problems.
Performance should be defined in a broader sense that includes not only recovery accuracy,
but also other properties of the algorithm, such as convergence speed.
Generally, for a nonconvex regularized algorithm, the performance is
usually closely related to the initialization and the convergence rate
is usually slower than that of a convex regularized algorithm.
In comparison, a convex algorithm has better stability and convergence properties,
and is insensitive to initialization since converging to a global minimal
is usually easy guaranteed [202]. Meanwhile, for first-order convex algorithms,
well-developed acceleration techniques can be used with guaranteed convergence [26].
However, when such acceleration techniques are applied to nonconvex algorithms,
there is no guarantee of the convergence and stability.

Therefore, the question of whether to use convex or nonconvex regularization
models requires careful deliberation. We show that convex models may be preferable
when the signal is not strictly sparse (or the matrix is not strictly low-rank) or
the signal-to-noise ratio (SNR) is low, since in these cases the performance improvement
of nonconvex models are often not distinct and may not deserve the price of more slowly
converging nonconvex algorithms. We provide a number of concrete examples that clarify these points.

We also address the issues of penalty selection and the convergence of related
nonconvex algorithms. We hope that our paper will illuminate the role nonconvex
regularization plays in sparse and low-rank recovery in signal processing,
statistics and machine learning, and demonstrate when and how it should be used.

There also exist some recent overview articles related to the topics of this work, e.g.,
on low-rank matrix recovery [203], [204], and nonconvex optimization in machine leaning [205].
While the works [203], [204] mainly focus on matrix factorization based (nonconvex) low-rank recovery problems,
the low-rank recovery problems investigated in this work are more general.
Meanwhile, while the article [205] mainly focuses on the optimization aspect of
general nonconvex problems in machine learning,
we focus on nonconvex regularized problems characterized with the
non-convexity and non-smoothness of the involved problems.
Moreover, we provide a more wide scope on the applications in various fields.

\textit{Outline}: The rest of this paper is organized as follows.
In section II, we review the
proximity operator for nonconvex penalties, and present extended vector proximity operator
(for joint sparse recovery) and singular value shrinkage operator (for low-rank matrix recovery)
for generalized nonconvex penalties.
Section III discusses nonconvex regularized sparse vector recovery problems,
including CS, sparse regression, sparse signals separation and sparse PCA.
Section IV reviews nonconvex regularized sparse matrix recovery problems,
including large sparse covariance and inverse covariance matrices estimation.
Section V discusses nonconvex regularized low-rank recovery problems,
including matrix completion and robust PCA.
Section VI further discusses other
applications involving nonconvex regularized sparse and low-rank recovery.
Section VII concludes the overview.

\textit{Notations}: For a matrix ${\mathbf{M}}$, ${\rm{rank}}({\mathbf{M}})$,
${\rm{tr}}({\mathbf{M}})$, $|{\mathbf{M}}|$, ${\left\| {\mathbf{M}} \right\|_2}$
and ${\left\| {\mathbf{M}} \right\|_{\rm{F}}}$ are the rank, trace, determinant,
spectral norm and Frobenius norm, respectively, whilst ${\rm{ei}}{{\rm{g}}_{\max }}({\mathbf{M}})$,
${\rm{ei}}{{\rm{g}}_{\min }}({\mathbf{M}})$ and ${\sigma _i}({\mathbf{M}})$ denote the maximal eigenvalue,
minimal eigenvalue and  the $i$-th largest singular value of ${\mathbf{M}}$.
For a matrix ${\mathbf{M}}$, ${\rm{diag}}({\mathbf{M}})$ is a diagonal matrix
which has the same diagonal elements as that of ${\mathbf{M}}$,
whilst for a vector ${\mathbf{v}}$, ${\rm{diag}}({\mathbf{v}})$ is a diagonal matrix
with diagonal elements be ${\mathbf{v}}$.
${\mathbf{M}} \ge {\mathbf{0}}$ mean that ${\mathbf{M}}$ is positive-semidefinite.
$\left\langle { \cdot , \cdot } \right\rangle$ and ${( \cdot )^T}$
stand for the inner product and transpose, respectively.
$\nabla f$ and $\partial f$ stand for the gradient and
subdifferential of the function $f$, respectively.
${\rm{sign}}( \cdot )$ denotes the sign of a quantity with ${\rm{sign}}(0){\rm{ = }}0$.
${{\mathbf{I}}_m}$ stands for an $m\times m$ identity matrix.
${\| \cdot \|_q}$ with $q \geq 0$ denotes the $\ell_q$-norm defined as
${\left\| {\mathbf{x}} \right\|_q} = {(\sum\nolimits_{i} {{{\left| {{x_i}} \right|}^q}} )^{1/q}}$.
${\delta _{i,j}}$ is the Kronecker delta function. $E\{\cdot\}$ denotes the expectation.
$I(\cdot)$ denotes the indicator function.

\section{Proximity Operator for Nonconvex Regularization Penalties}

Proximity operator plays a central role in developing efficient
proximal splitting algorithms for many optimization problems,
especially for nonconvex and nonsmooth inverse problems
encountered in the applications addressed in this paper.
As will be shown later, for convex or nonconvex penalized minimization problems,
proximity operator is the core of most highly-efficient first-order
algorithms which scale well for high-dimensional problems.
This section reviews nonconvex regularization penalties and
their corresponding proximity operator, including the hard-thresholding, $\ell_q$-norm,
an explicit $q$-shrinkage, SCAD, MCP, and firm thresholding.

\subsection{Scalar Proximity Operator}

For a proper and lower semi-continuous penalty function ${P_\lambda }( \cdot )$
where $\lambda > 0$ is a threshold parameter, consider the following scalar proximal mapping
\begin{equation}
{\rm{pro}}{{\rm{x}}_{{P_\lambda }}}(t) = \arg \mathop {\min }\limits_x \left\{ {{P_\lambda }(x) + \frac{1}{2}{{(x - t)}^{\rm{2}}}} \right\}.
\end{equation}
As ${P_\lambda }( \cdot )$ is separable, the proximity operator of a vector
${\mathbf{t}} = {[{t_1}, \cdots ,{t_n}]^T} \in {\mathbb{R}^n}$, denoted by ${\rm{pro}}{{\rm{x}}_{{P_\lambda }}}({\mathbf{t}})$,
can be computed in an element-wise manner as
\begin{equation}
{\rm{pro}}{{\rm{x}}_{{P_\lambda }}}({\mathbf{t}}) = {[{\rm{pro}}{{\rm{x}}_{{P_\lambda }}}({t_1}), \cdots ,{\rm{pro}}{{\rm{x}}_{{P_\lambda }}}({t_n})]^T}.
\end{equation}
Table 1 shows the penalties and the corresponding proximal mapping operator.
Among the presented penalties, only the soft-thresholding penalty is convex,
while the other penalties are (symmetric) folded concave functions, as shown in Fig. 1.

The hard-thresholding was first introduced in [36] and then applied for wavelet
applications in statistics [4], which is a natural selection for sparse inducing [77]--[81].
The well-known soft-thresholding rule was first observed by Donoho, Johnstone,
Hoch and Stem [35] and then used in wavelet applications [36], which forms
the core of the LASSO introduced by Tibshirani [37].
The $\ell_1$ penalty is the most popular one as its convexity makes
the related optimization problems more tractable than that using a nonconvex one.

However, the $\ell_1$ penalty has a bias problem. More specifically,
when the true parameter has a relatively large magnitude, the
soft-thresholding estimator is biased since it imposes a constant
shrinkage on the parameter, as shown in Fig. 2. Fig. 2 plots the
thesholding/shrinkage functions for the hard-, soft-, $\ell_q$- and
SCAD penalties with a same threshold. In contrast, the hard-thresholding
and SCAD estimators are unbiased for large parameter. Meanwhile,
the thresholding rule corresponding to SCAD, $\ell_q$ and $q$-shrinkage
fall in (sandwiched) between hard- and soft-thresholding.

\begin{figure}[!t]
 \centering
 \includegraphics[scale = 0.68]{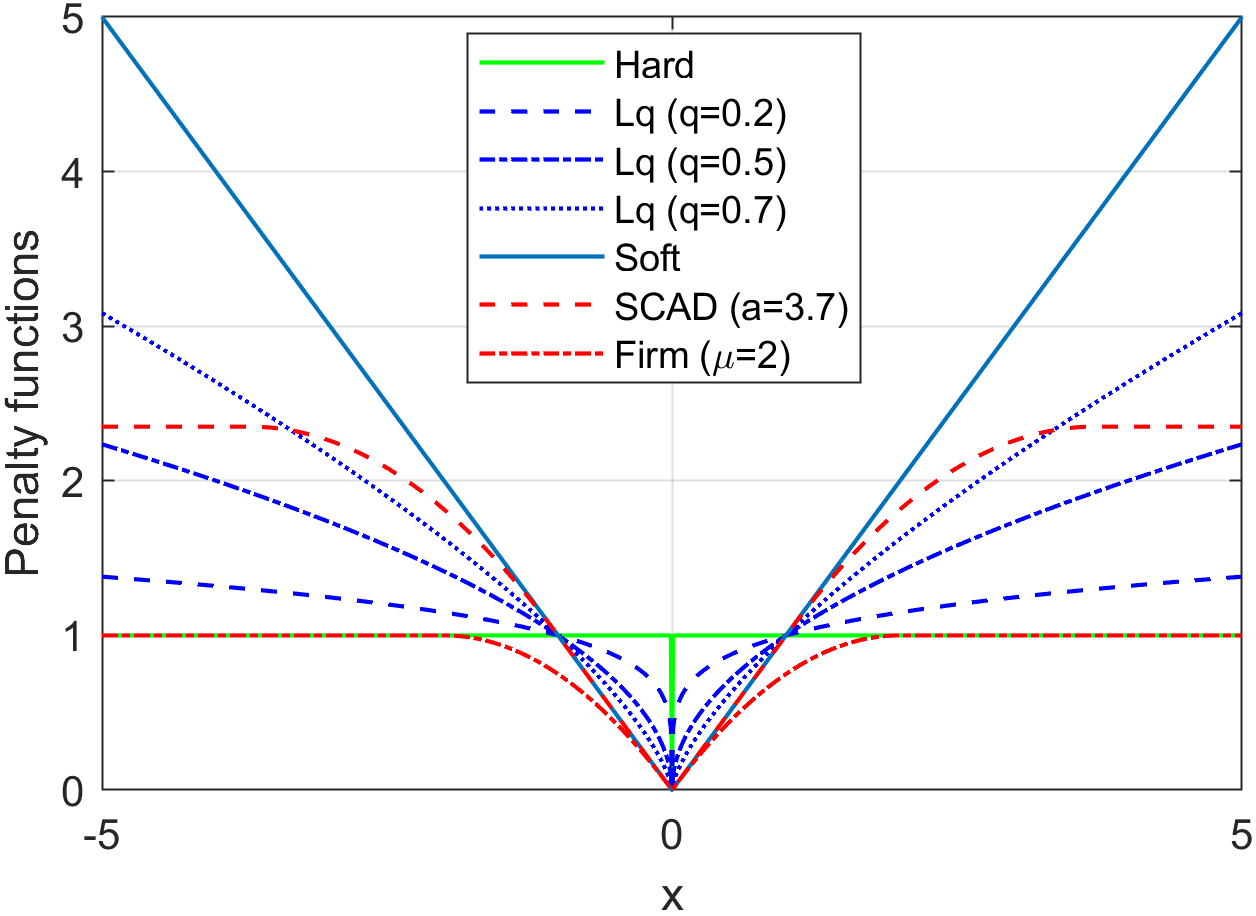}
\caption{Plot of penalty functions for $\lambda = 1$.}
 \label{figure1}
\end{figure}

\begin{figure}[!t]
 \centering
 \includegraphics[scale = 0.68]{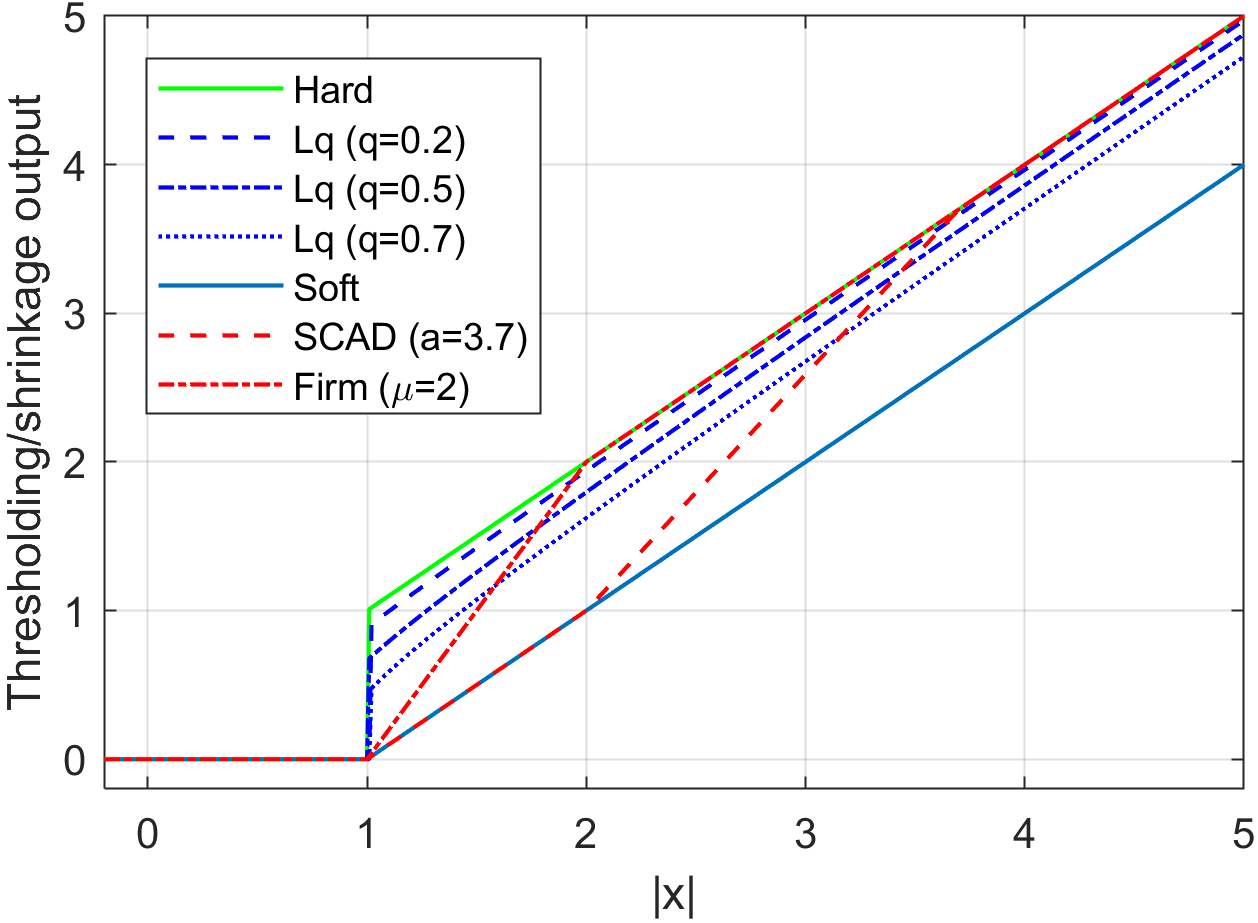}
\caption{Plot of thesholding/shrinkage functions (with a same threshold).}
 \label{figure2}
\end{figure}

\renewcommand\arraystretch{1.5}
\begin{table*}[!t]
\caption{Regularization penalties and the corresponding proximity operator ($\lambda > 0$ is a thresholding parameter).}
\footnotesize
\centering
\newcommand{\tabincell}[2]{\begin{tabular}{@{}#1@{}}#2\end{tabular}}
\begin{tabular}{|p{3.5cm}|p{6cm}|p{7cm}|}
\hline
Penalty name & Penalty formulation & Proximity operator \\
\hline
(i)  Soft thresholding [35] & ${P_\lambda }(x) = \lambda {\rm{|}}x{\rm{|}}$ & ${\rm{pro}}{{\rm{x}}_{{P_\lambda }}}(t) = {\rm{sign}}(t)\max \left\{ {|t| - \lambda,0} \right\}$ \\

\hline
(ii) Hard thresholding [36] & \tabincell{c}{${P_\lambda }(x) = \lambda [2 - {({\rm{|}}x{\rm{|}} - \sqrt 2 )^2}I({\rm{|}}x{\rm{|}} < \sqrt 2 )]$\\ or ${P_\lambda }(x) = \lambda {\rm{|}}x{{\rm{|}}_0}$} & ${\rm{pro}}{{\rm{x}}_{{P_\lambda }}}(t) = \left\{ {\begin{array}{*{20}{l}}
{0,}&{|t| < \sqrt {2\lambda } }\\
{\{ 0,t\} ,}&{|t| = \sqrt {2\lambda } }\\
{t,}&{|t| > \sqrt {2\lambda } }
\end{array}} \right.$ \\

\hline
(iii) $\ell_q$-norm [6], [7]  & ${P_\lambda }(x) = \lambda {\rm{|}}x{{\rm{|}}^q}$, $0 < q < 1$  &
\tabincell{c}{${\rm{pro}}{{\rm{x}}_{{P_\lambda }}}(t) = \left\{ {\begin{array}{*{20}{l}}
{0,}&{{\rm{|}}t{\rm{|}} < \tau }\\
{\{ 0,{\rm{sign}}(t)\beta \} ,}&{{\rm{|}}t{\rm{|}} = \tau }\\
{{\rm{sign}}(t)y,}&{{\rm{|}}t{\rm{|}} > \tau }
\end{array}} \right.$\\
where $\beta  = {[2\lambda (1 - q)]^{1/(2 - q)}}$, $\tau = \beta  + \lambda q{\beta ^{q - 1}}$,\\
 $h(y) = \lambda q{y^{q - 1}} + y - \left| t \right| = 0$ and $y \in [\beta ,{\rm{|}}t{\rm{|}}]$}\\

\hline
(iv) $q$-shrinkage [10]  & N/A ($q < 1$) &  ${\rm{pro}}{{\rm{x}}_{{P_\lambda }}}(t) = {\rm{sign}}(t)\max \left\{ {|t| - {\lambda ^{2 - q}}{t^{q - 1}},0} \right\}$\\

\hline
(v)  SCAD [11]  &  ${P_\lambda }(x) = \left\{ {\begin{array}{*{20}{l}}
{\lambda {\rm{|}}x{\rm{|}},}&{{\rm{|}}x{\rm{|}} < \lambda }\\
{\frac{{2a\lambda {\rm{|}}x{\rm{|}} - {x^2} - {\lambda ^2}}}{{2(a - 1)}},}&{\lambda  \le {\rm{|}}x{\rm{|}} < a\lambda }\\
{(a + 1){\lambda ^2}/2,}&{{\rm{|}}x{\rm{|}} \ge a\lambda }
\end{array}} \right.$  &  ${\rm{pro}}{{\rm{x}}_{{P_\lambda }}}(t) = \left\{ {\begin{array}{*{20}{l}}
{{\rm{sign}}(t)\max \{ {\rm{|}}t{\rm{|}} - \lambda ,0\} ,}&{{\rm{|}}t{\rm{|}} \le 2\lambda }\\
{\frac{{(a - 1)t - {\rm{sign}}(t)a\lambda }}{{a - 2}},}&{2\lambda  < {\rm{|}}t{\rm{|}} \le a\lambda }\\
{t,}&{{\rm{|}}t{\rm{|}} > a\lambda }
\end{array}} \right.$\\

\hline
(vi) MCP [12]   &  \tabincell{c}{${P_{\lambda ,\gamma }}(x) = \lambda \int_0^{|x|} {\max (1 - t/(\gamma \lambda ),0)} dt$,\\ where $\gamma  > 0$}   &
${\rm{pro}}{{\rm{x}}_{{P_{\lambda ,\gamma }}}}(t) = \left\{ {\begin{array}{*{20}{l}}
{0,}&{{\rm{|}}t{\rm{|}} \le \lambda }\\
{\frac{{{\rm{sign}}(t)(|t| - \lambda )}}{{1 - 1/\gamma }},}&{\lambda  < {\rm{|}}t{\rm{|}} \le \gamma \lambda }\\
{t,}&{{\rm{|}}t{\rm{|}} > \gamma \lambda }
\end{array}} \right.$\\

\hline
(vii) Firm thresholding [13]  &  ${P_{\lambda ,\mu }}(x) = \left\{ {\begin{array}{*{20}{l}}
{\lambda [|x| - {x^2}/(2\mu )],}&{|t| \le \mu }\\
{\lambda \mu /2,}&{|t| \ge \mu }
\end{array}} \right.$  where $\mu > \lambda$   &  ${\rm{pro}}{{\rm{x}}_{{P_{\lambda ,\mu }}}}(t) = \left\{ {\begin{array}{*{20}{l}}
{0,}&{{\rm{|}}t{\rm{|}} \le \lambda }\\
{\frac{{{\rm{sign}}(t)(|t| - \lambda )\mu }}{{\mu  - \lambda }},}&{\lambda  \le {\rm{|}}t{\rm{|}} \le \mu }\\
{t,}&{{\rm{|}}t{\rm{|}} \ge \mu }
\end{array}} \right.$\\
\hline
\end{tabular}
\end{table*}
\renewcommand\arraystretch{1}

The ${\ell_q}$ penalty with $0<q<1$ bridges the gap between the
${\ell_0}$ and ${\ell_1}$ penalties, and intuitively its shrinkage
function is less biased than soft-thresholding. The proximity operator
of the ${\ell _q}$-norm penalty does not has a closed-form expression
except for the two special cases of $q = 1/2$ and $q = 2/3$ [5]
(in these two cases the proximal mapping can be explicitly expressed
as the solution of a cubic or quartic equation), but it can be
efficiently solved, e.g., by a Newton's method. Moreover,
explicit $q$-shrinkage mapping has been proposed in [8]--[10],
which has some qualitative resemblance to the $\ell_q$ proximal mapping
while being continuous and explicit. The $q$-shrinkage reduces to the
soft-thresholding when $q=1$, while it tends pointwise to the hard-thresholding
in the limit as $q \to  - \infty$. As an acceptable price to pay for
having an explicit proximal mapping, its penalty does not have a closed-form expression.

The SCAD penalty has been widely used in variable selection problems,
and it has shown favorable effectiveness compared with other penalties
in high-dimensional variable selection problems [11].
As well as the $\ell_0$, $\ell_q$, and SCAD penalties,
the MCP penalty can also ameliorate the bias problem of the $\ell_1$ penalty [12],
and it has been widely used for penalized variable selection in
high-dimensional linear regression. For the MCP penalty with each $\lambda > 0$,
we can obtain a continuum of penalties and threshold operators by varying
$\gamma$ in the range $(0, + \infty )$. Moreover, the firm thresholding
is a continuous and piecewise-linear approximation of the hard-thresholding [13].
In addition, a class of nonconvex penalties which can maintain the convexity of
the global cost function have been designed in [209]--[211],
whilst log penalties have been considered in [212], [213].

Among the presented shrinkage functions, the soft-, SCAD, $q$-shrinkage and
firm thresholding are continuous while the hard- and ${\ell _q}$-thresholding
are discontinuous. For each of the presented penalties,
the corresponding thresholding/shrinkage operator ${\rm{pro}}{{\rm{x}}_{{P_\lambda }}}(t)$ satisfies
\begin{equation}
\begin{split}
& \mathrm{i)}~~~ {\rm{pro}}{{\rm{x}}_{{P_\lambda }}}(t) = {\rm{sign}}(x){\rm{pro}}{{\rm{x}}_{{P_\lambda }}}(|t|)\\
& \mathrm{ii)}~~ \left| {{\rm{pro}}{{\rm{x}}_{{P_\lambda }}}(t)} \right| \le |t|\\
& \mathrm{iii)}~~ {\rm{pro}}{{\rm{x}}_{{P_\lambda }}}(t) = 0~~ \mathrm{for~ some ~threshold}~~ |t| \le {T_\lambda }\\
& \mathrm{iv)}~\left| {{\rm{pro}}{{\rm{x}}_{{P_\lambda }}}(t) - t} \right| \le \lambda
\end{split}
\end{equation}
where ${T_\lambda } > 0$ is a threshold dependent on $\lambda $.
These conditions establish the sign consistency, shrinkage, thresholding,
and limited shrinkage properties for the thresholding/shrinkage operator corresponding to a generalized penalty.

\subsection{Vector Proximity Operator (for Multitask Joint Sparse Recovery)}

In many applications, as will be shown in section III,
it is desirable to jointly recover multichannel signals to exploit
the multichannel joint sparsity pattern.
In this case, it involves solving the following vector optimization problem
\begin{equation}
{\mathbf{pro}}{{\mathbf{x}}_{{P_\lambda }}}({\mathbf{t}}) =
\arg \mathop {\min }\limits_{\mathbf{x}} \left\{ {{P_\lambda }({{\left\| {\mathbf{x}} \right\|}_2}) + \frac{1}{2}\left\| {{\mathbf{x}} - {\mathbf{t}}} \right\|_2^2} \right\}
\end{equation}
where ${\mathbf{x}} \in {\mathbb{R}^L}$ and ${\mathbf{t}} \in {\mathbb{R}^L}$ with $L$ be the channel number.
The case of ${P_\lambda }$ be the ${\ell _q}$ penalty has been considered in [120],
we extend the result to generalized penalties in the following.

\textbf{Theorem 1}. Suppose that ${P_\lambda }(x)$ is a nondecreasing
function for $x \in [0,\infty )$, for any ${\mathbf{x}} \in {\mathbb{R}^L}$,
the solution to (4) is given by
\begin{equation}
{\mathbf{pro}}{{\mathbf{x}}_{{P_\lambda }}}({\mathbf{t}}) = \left\{ {\begin{array}{*{20}{l}}
{{\mathbf{0}},}&{{\mathbf{t}} = {\mathbf{0}}}\\
{\frac{{{\rm{pro}}{{\rm{x}}_{{P_\lambda }}}({{\left\| {\mathbf{t}} \right\|}_2})}}{{{{\left\| {\mathbf{t}} \right\|}_2}}}{\mathbf{t}},}&{{\rm{otherwise}}}
\end{array}} \right. .
\end{equation}

\textit{Proof}: See Appendix A.

\subsection{Singular Value Shrinkage Operator (for Low-Rank Matrix Recovery)}

For a matrix ${\mathbf{M}} \in {\mathbb{R}^{m \times n}}$,
consider the singular value decomposition (SVD) of rank $r$, ${\mathbf{M}} = {\mathbf{U\Sigma }}{{\mathbf{V}}^T}$,
where ${\mathbf{\Sigma }} = {\rm{diag}}\{ {\sigma _1}, \cdots ,{\sigma _r}\}$
contains the singular values, ${\mathbf{U}} \in {\mathbb{R}^{m \times r}}$ and
${\mathbf{V}} \in {\mathbb{R}^{r \times n}}$ contain the orthonormal singular vectors.
In low-rank matrix recovery, the low-rankness promotion on a matrix is usually
achieved by sparsity promotion on the sigular values of the matrix.
We denote a generalized penalty for low-rankness promotion by ${\bar P_\lambda }$,
which is defined as
\begin{equation} 
{\bar P_\lambda }({\mathbf{M}}) = \sum\nolimits_i {{P_\lambda }({\sigma _i})}
\end{equation}
where ${P_\lambda }$ is a generalized penalty for sparsity inducing as introduced in Table 1.
In the two cases of ${P_\lambda }(\cdot) = {\left\|  \cdot\right\|_0}$ and
${P_\lambda }(\cdot) = {\left\|  \cdot\right\|_1}$, ${\bar P_\lambda }({\mathbf{M}})$
becomes the rank ${\rm{rank}}({\mathbf{M}})$ and nuclear norm ${\left\| {\mathbf{M}} \right\|_*}$ of ${\mathbf{M}}$, respectively.
When ${P_\lambda }$ is the ${\ell _q}$ penalty, ${\bar P_\lambda }$
becomes the Schatten-$q$ quasi-norm of matrix.

In the following, we provide the generalized singular-value shrinkage operator
for a generalized penalty ${\bar P_\lambda }$.

\textbf{Theorem 2}. For a rank $r$ matrix ${\mathbf{M}} \in {\mathbb{R}^{m \times n}}$,
suppose that it has an SVD ${\mathbf{M}} = {\mathbf{U\Sigma }}{{\mathbf{V}}^T}$,
where ${\mathbf{\Sigma }} = {\rm{diag}}\{ {\sigma _1}, \cdots ,{\sigma _r}\}$ and
${\sigma _1} \ge {\sigma _2} \cdots  \ge {\sigma _r}$, ${\mathbf{U}}$ and ${\mathbf{V}}$
contain the left and right singular vectors. Then, for any ${\bar P_\lambda }$ defined
as (6) with ${P_\lambda}$ satisfying (3),
the solution to the optimization problem
\begin{equation} 
{\rm{pro}}{{\rm{x}}_{{{\bar P}_\lambda }}}({\mathbf{M}}) = \arg \mathop {\min }\limits_{\mathbf{X}} \left\{ {{{\bar P}_\lambda }({\mathbf{X}}) + \frac{1}{2}\left\| {{\mathbf{X}} - {\mathbf{M}}} \right\|_{\rm{F}}^2} \right\}
\end{equation}
is given by
\begin{equation} 
{\rm{pro}}{{\rm{x}}_{{{\bar P}_\lambda }}}({\mathbf{M}}) = {\mathbf{U}} \cdot {\rm{diag}}\left\{ {{\rm{pro}}{{\rm{x}}_{{P_\lambda }}}({\sigma _1}, \cdots ,{\sigma _r})} \right\} \cdot {{\mathbf{V}}^T}
\end{equation}
where ${\rm{pro}}{{\rm{x}}_{{P_\lambda }}}$ is the proximity operator defined in (2).

\textit{Proof}: See Appendix B.

\section{Sparse Vector Recovery}

This section reviews nonconvex regularization based sparse vector signals recovery,
mainly on the following four topics, CS, sparse regression and variable selection,
sparse signals separation with application to image inpainting and super-resolution,
and sparse PCA. Strictly speaking, sparse PCA is not a vector recovery problem,
but in many popular greedy approaches, the principle components are estimated in
a one-by-one (vector-by-vector) manner.

\subsection{Compressive Sensing}

In the past decade, compressive sensing has attracted extensive studies [14]--[17]
and has found wide applications in radar [18], [19], communications [20], medical imaging [21],
image processing [22], and speech signal processing [23]. In the CS framework,
sparse signals (or signals can be sparsely represented in some basis) can be acquired
at a significantly lower rate than the classical Nyquist sampling, and signals only need
to be sampled at a rate proportional to their information content.

In CS, the objective is to reconstruct a sparse signal ${\mathbf{x}} \in {\mathbb{R}^n}$ from
its compressed measurement
\begin{equation} 
{\mathbf{y}} = {\mathbf{Ax}} + {\mathbf{n}}
\end{equation}
where ${\mathbf{A}} \in {\mathbb{R}^{m \times n}}$ with $m < n$ is the sensing matrix
(also called measurement matrix), ${\mathbf{n}} \in {\mathbb{R}^m}$ is additive measurement noise.
Since $m < n$, the recovery of ${\mathbf{x}}$ from the compressed measurement is generally ill-posed.
However, provided that ${\mathbf{x}}$ is sparse and the sensing matrix ${\mathbf{A}}$ satisfies some
stable embedding conditions [17], ${\mathbf{x}}$ can be reliably recovered with an error upper
bounded by the noise strength. This can be achieved in the noiseless case by the formulation
\begin{equation} 
\begin{split}
&\mathop {{\mathrm{min}}}\limits_{\mathbf{x}} P({\mathbf{x}})\\
\textrm{subject~ to}&~~ {\mathbf{Ax}} = {\mathbf{y}}~~~~~~~~~~~~~
\end{split}
\end{equation}
or in the noisy case by the formulation
\begin{equation} 
\begin{split}
&\mathop {{\mathrm{min}}}\limits_{\mathbf{x}} P({\mathbf{x}})\\
\textrm{subject~ to}&~~ {\left\| {{\mathbf{Ax}} - {\mathbf{y}}} \right\|_2} \le \sigma
\end{split}
\end{equation}
where $P$ is a penalty for sparsity inducing (a special case of ${P_\lambda }$ with $\lambda = 1$),
and $\sigma > 0$ bounds the ${\ell_2}$-norm of the residual error.
This constrained formulation (11) can be converted into an unconstrained formulation as
\begin{equation} 
\mathop {{\mathrm{min}}}\limits_{\mathbf{x}} \frac{1}{2}\left\| {{\mathbf{Ax}} - {\mathbf{y}}} \right\|_2^2 + {P_\lambda }({\mathbf{x}}).
\end{equation}

Naturally, using the ${\ell_0}$-norm penalty, i.e., $P({\mathbf{x}}) = {\left\| {\mathbf{x}} \right\|_0}$
and ${P_\lambda }({\mathbf{x}}) = \lambda {\left\| {\mathbf{x}} \right\|_0}$,
which counts the number of nonzero components in the vector ${\mathbf{x}}$,
(10), (11) and (12) are the exact formulations of finding a sparse vector
to fulfill the linear constraint ${\mathbf{Ax}} = {\mathbf{y}}$, satisfy the residual
constraint ${\left\| {{\mathbf{Ax}} - {\mathbf{y}}} \right\|_2} \le \sigma$,
and minimize the quadratic loss function in (12), respectively.
However, with the ${\ell _0}$ penalty the problems (10)-(12) are nonconvex and NP-hard,
thus, convex relaxation methods are often considered, e.g.,
replace the ${\ell _0}$ penalty by the ${\ell _1}$ one.
With $P({\mathbf{x}}) = {\left\| {\mathbf{x}} \right\|_1}$ and ${P_\lambda }({\mathbf{x}}) = \lambda {\left\| {\mathbf{x}} \right\|_1}$,
the formulations (10), (11) and (12) are the well-known basis-pursuit (BP) [24],
basis-pursuit denoising (BPDN) and LASSO [13], respectively.
In this case, the formulations are convex and hence tractable.
A large number of algorithms have been developed in the past decade
for these $\ell_1$ minimization problems (see [25]--[29] and the reference therein).

The CS theory has established that if the sensing matrix satisfies some conditions,
such as the restricted isometry property (RIP) [17], [30]--[32],
the null space property [33], and the incoherence condition [34],
the sparse signal can be reconstructed by $\ell_1$ regularization reliably.
However, due to the relaxation, the recovery accuracy is often degraded, e.g.,
it often introduces extra bias [1], [2] and cannot reconstruct a signal
with the least observations [3]. Furthermore, for some applications,
the result of the $\ell_1$-minimization is not sparse enough and the original signals cannot be recovered.
A simple example of such a case has been given in [38] with intuitive explanation.

To address this problem, a number of improved algorithms have been developed via
employing the nonconvex $\ell_q$-norm penalty instead of the $\ell_1$ one, i.e.,
$P({\mathbf{x}}) = {\left\| {\mathbf{x}} \right\|_q}$ and ${P_\lambda }({\mathbf{x}}) = \lambda \left\| {\mathbf{x}} \right\|_q^q$ with $0 < q < 1$.
For $0 < q < 1$, $\left\| {\mathbf{x}} \right\|_q^q$ is the ${\ell _q}$ quasi-norm
defined as $\left\| {\mathbf{x}} \right\|_q^q = \sum\nolimits_i {{{\left| {{x_i}} \right|}^q}}$.
Compared with the ${\ell _1}$-norm, the ${\ell _q}$-norm is a closer approximation of the ${\ell _0}$-norm.
It has been shown in [37] that under certain RIP conditions of the sensing matrix,
${\ell _q}$-regularized algorithms require fewer measurements to achieve reliable
recovery than ${\ell _1}$-regularized algorithms. Moreover, the sufficient conditions
in terms of RIP for ${\ell _q}$ regularization are weaker than those for ${\ell _1}$ regularization [39], [40].
Meanwhile, it has been shown in [41] that for any given measurement matrix with
restricted isometry constant ${\delta _{2k}} < 1$, there exists some $q \in (0,1)$ that
guarantees exact recovery of signals with support smaller than $k < m/2$ by $\ell_q$-minimization.

Recently, $\ell_q$-regularized sparse reconstruction has been extensively studied for CS,
e.g., [5], [8], [10], [39]--[66], [200], and extended to structured sparse recovery [201].
As the $\ell_q$ penalty is nonsmooth and nonconvex, many of these algorithms solve a smoothed
(approximated) $\ell_q$-minimization problem, e.g., the works [47]--[49] use an approximation
of $\left\| {\mathbf{x}} \right\|_q^q$ as
\begin{equation} 
\left\| {\mathbf{x}} \right\|_{q,\varepsilon }^q = \sum\limits_{i = 1}^n {{{\left( {x_i^2 + {\varepsilon ^2}} \right)}^{\frac{q}{2}}}}
\end{equation}
where $\varepsilon>0$ is a smoothing parameter.
Furthermore, the iteratively reweighted algorithms [38], [44], [46] use the following two penalties (at the $k+1$-th iteration)
\begin{equation} 
\begin{split}
&\left\| {{{\mathbf{x}}}} \right\|_{q,\varepsilon }^q = \sum\limits_{i = 1}^n {{{\left( {\left| {x_i^k} \right| + \varepsilon } \right)}^{q - 1}}\left| {{x_i}} \right|}\\
&\left\| {{{\mathbf{x}}}} \right\|_{q,\varepsilon }^q = \sum\limits_{i = 1}^n {{{\left( {{{\left| {x_i^k} \right|}^2} + {\varepsilon ^2}} \right)}^{\frac{q}{2} - 1}}{{\left| {{x_i}} \right|}^2}}  \notag
\end{split}
\end{equation}
which explicitly relate to the $\ell_q$-norm approximation.

These algorithms have been shown to achieve better recovery performance relative to
the $\ell_1$-regularized algorithms. However, due to the non-convexity of $\ell_q$-minimization,
many of these algorithms are generally inefficient and impractical for large-scale problems.
For example, the StSALq method in [47] requires repetitive computation of matrix inversion of
dimension $n \times n$, whilst the IRucLq method in [48] solve a set of linear equations using
matrix factorization (with matrix dimension of $n \times n$). In comparison, the Lp-RLS method in [49],
which uses an efficient conjugate gradient (CG) method in solving a sequence of smoothed subproblems,
has better scalable capability. Moreover, the iteratively reweighted method in [50] also involves
solving a sequence of weighted $\ell_1-\ell_2$ mixed subproblems. Although efficient first-order
algorithms can be used to solve the subproblems involved in the methods in [49], [50],
both the methods have a double loop which hinders the overall efficiency.
While subsequence convergence is guaranteed for the iteratively reweighted methods [48] and [50],
there is no such guarantee for StSALq [47] and Lp-RLS [49].

In comparison, the proximal gradient descent (PGD) and
alternative direction method of multipliers (ADMM)
algorithms for the problem (12) are globally convergent
under some mild conditions, while being much more efficient.
Specifically, let $f({\mathbf{x}}) = 1/2\left\| {{\mathbf{Ax}} - {\mathbf{y}}} \right\|_2^2$,
consider the following quadratic approximation of the objective in the formulation (12) at iteration $k+1$
and at a given point ${{\mathbf{x}}^k}$ as
\begin{equation}  
\begin{split}
&{Q_{{L_f}}}({\mathbf{x}};{{\mathbf{x}}^k})\\
&= f({{\mathbf{x}}^k}) + {\left( {{\mathbf{x}} - {{\mathbf{x}}^k}} \right)^T}\nabla f({{\mathbf{x}}^k}) + \frac{{{L_f}}}{2}\left\| {{\mathbf{x}} - {{\mathbf{x}}^k}} \right\|_2^2 + {P_\lambda }({\mathbf{x}})
\end{split}
\end{equation}
where ${L_f} > 0$ is a proximal parameter. Then, minimizing ${Q_{{L_f}}}({\mathbf{x}};{{\mathbf{x}}^k})$
reduces to the proximity operator introduced in section 2 as
\begin{equation}  
{{\mathbf{x}}^{k + 1}} = {\mathrm{pro}}{{\mathrm{x}}_{(1/{L_f}){P_\lambda }}}\left( {{{\mathbf{x}}^k} - \frac{1}{{{L_f}}}\nabla f({{\mathbf{x}}^k})} \right)
\end{equation}
which can be computed element-wise via the shrinkage/thresholding function in Table I.

This PGD algorithm fits the frameworks of the forward-backward
splitting method [67] and the generalized gradient projection method [68].
Very recently, the convergence properties of this kind of algorithms have been established
via exploiting the Kurdyka-Lojasiewicz (KL) property of the objective function [69]--[71].
Suppose that ${P_\lambda }$ is a closed, proper, lower semi-continuous, KL function,
if ${L_f} > {\lambda _{\max }}({{\mathbf{A}}^T}{\mathbf{A}})$, then, the sequence $\{ {{\mathbf{x}}^k}\}$
generated by PGD (15) converges to a stationary point of the problem (12).
Further, under some more conditions, the convergence of PGD to
a local minimizer can be guaranteed [72], [73].

For ADMM algortithm, using an auxiliary variable
\[{\mathbf{z}} = {\mathbf{Ax}} - {\mathbf{y}} \]
the problem (12) can be equivalently reformulated as
\begin{equation}   
\begin{split}
&\mathop {{\mathrm{min}}}\limits_{\mathbf{x}} \frac{1}{2}\left\| {\mathbf{z}} \right\|_2^2 + {P_\lambda }({\mathbf{x}})\\
\textrm{subject~ to}&~~{\mathbf{Ax}} - {\mathbf{y}} - {\mathbf{z}} = {\mathbf{0}}.~~~~~~~~~~~~~~~~~~~~
\end{split}
\end{equation}
Then, the ADMM algorithm consists of the following steps
\begin{equation}
\begin{split}
{{\mathbf{x}}^{k + 1}} &= \arg \mathop {\min }\limits_{\mathbf{x}} \left( {{P_\lambda }({\mathbf{x}}) + \frac{\rho }{2}\left\| {{\mathbf{Ax}} - {\mathbf{y}} - {{\mathbf{z}}^k} + \frac{{{{\mathbf{w}}^k}}}{\rho }} \right\|_2^2} \right)\\
{{\mathbf{z}}^{k + 1}} &= \arg \mathop {\min }\limits_{\mathbf{z}} \left( {\frac{1}{2}\left\| {\mathbf{z}} \right\|_2^2 + \frac{\rho }{2}\left\| {{\mathbf{A}}{{\mathbf{x}}^{k + 1}} - {\mathbf{y}} - {\mathbf{z}} + \frac{{{{\mathbf{w}}^k}}}{\rho }} \right\|_2^2} \right)\\
{{\mathbf{w}}^{k + 1}} &= {{\mathbf{w}}^k} + \rho ({\mathbf{A}}{{\mathbf{x}}^{k + 1}} - {\mathbf{y}} - {{\mathbf{z}}^{k + 1}})    \notag
\end{split}
\end{equation}
where ${\mathbf{w}}$ is the dual variable, $\rho>0$ is a penalty parameter.
As a standard trick, the ${\mathbf{x}}$-subproblem can be solved approximately
via linearizaing the quadratic term. For a closed, proper, lower semi-continuous and KL ${P_\lambda }$,
under some condition of the proximal parameter and $\rho$ (should be chosen sufficiently large),
this proximal ADMM algorithm globally converges to a stationary point of the problem (16) [74]--[75].

For PGD and ADMM, the dominant computational load in each iteration is
matrix-vector multiplication with complexity $O(mn)$, thus,
scale well for high-dimension problems.
These two algorithms may be further accelerated by the schemes in [26], [76],
however, for a nonconvex ${P_\lambda}$, there is no guarantee of convergence
when using such acceleration schemes.

\begin{figure}[!t]
 \centering
 \includegraphics[scale = 0.65]{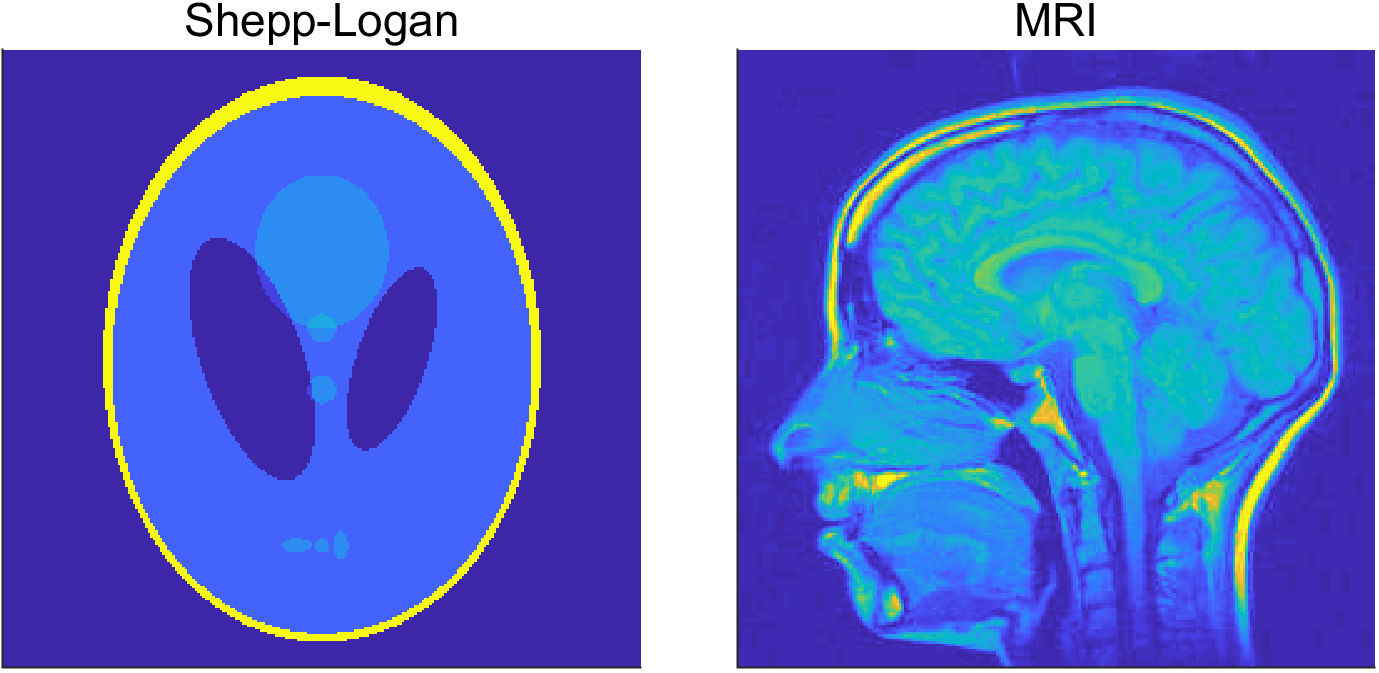}
\caption{The two considered $256 \times 256$ images.}
 \label{figure3}
\end{figure}

\begin{figure}[!t]
 \centering
 \includegraphics[scale = 0.65]{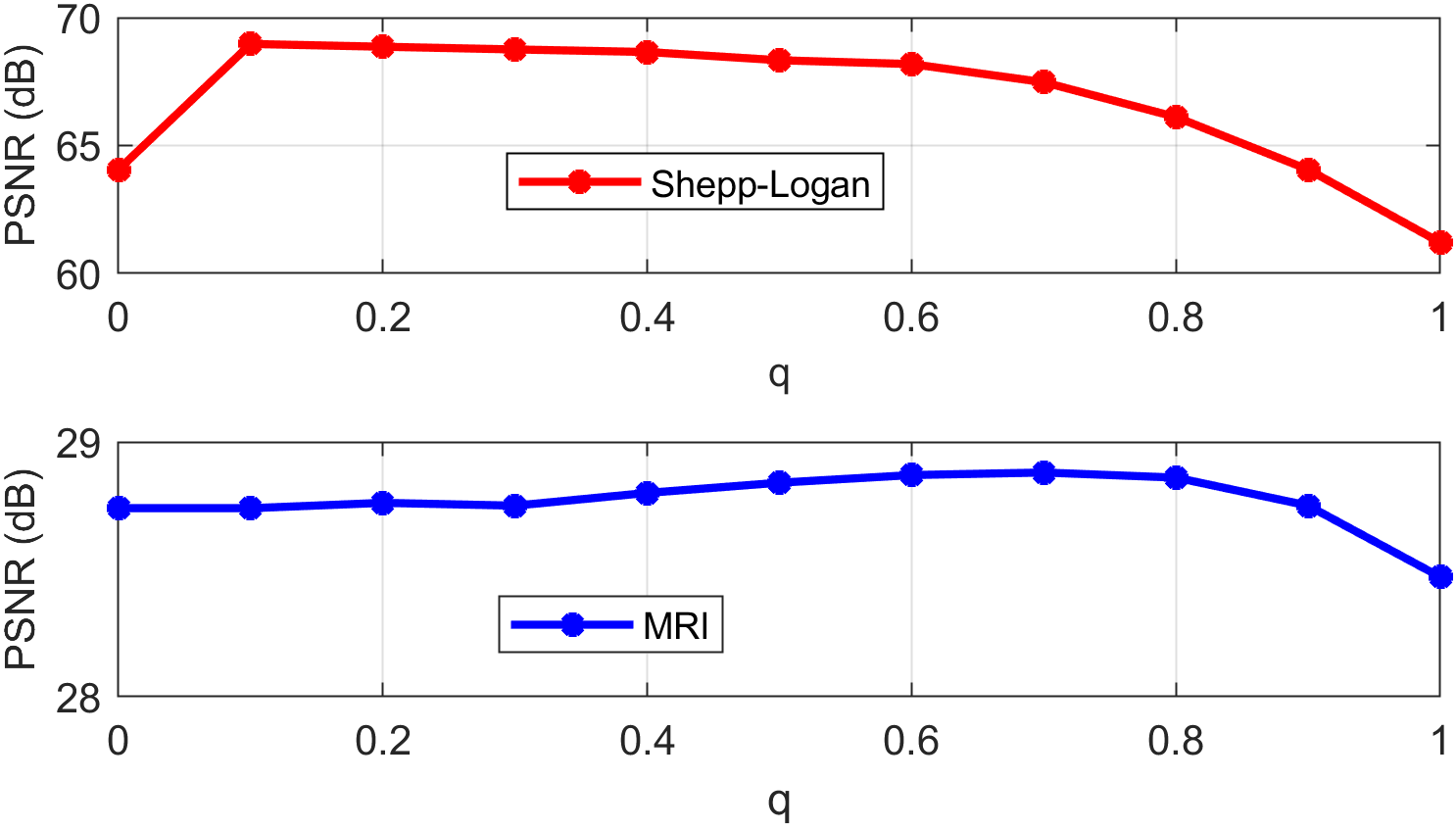}
\caption{Recovery performance of $\ell_q$-thresholding versus $q$ in reconstructing two $256 \times 256$ images.}
 \label{figure4a}
\end{figure}

\begin{figure}[!t]
 \centering
 \includegraphics[scale = 0.63]{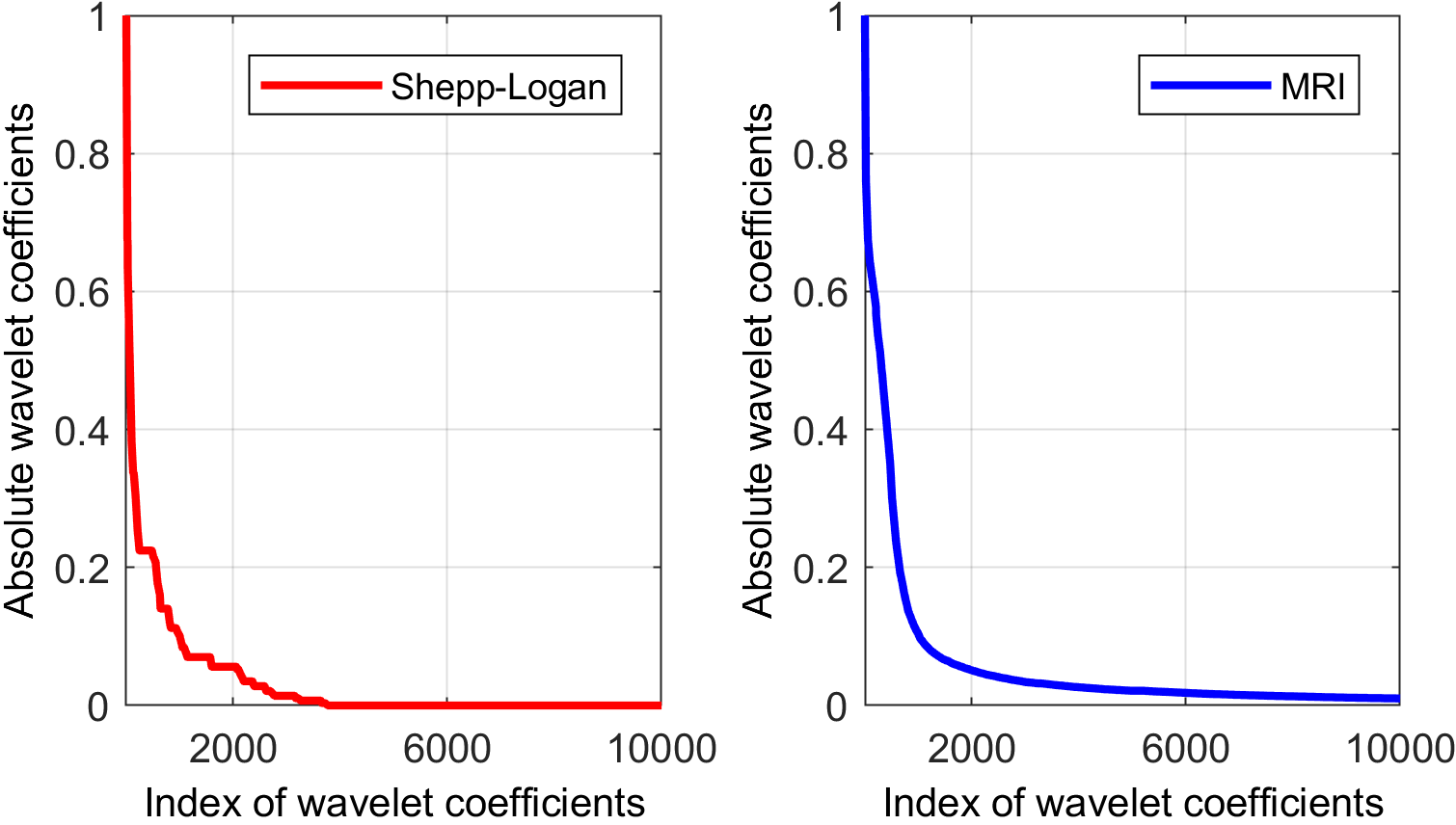}
\caption{Sorted and normalized absolute values of the wavelet coefficients of the two images (the first 10000 large values).}
 \label{figure4b}
\end{figure}

\textit{Example 1 (Image reconstruction)}.
We evaluate the PGD (15) with hard-, ${\ell _q}$- and soft-thresholding on image reconstruction.
The used images are of size $256 \times 256$ ($n = 65536$),
include a synthetic image, ``Shepp-Logan'' and an MRI image, as shown in Fig. 3.
We use the Haar wavelets as the basis for sparse representation of the images,
and ${\mathbf{A}}$ is a partial DCT matrix with $m = {\mathrm{round}}(0.4n)$.
The ${\ell _q}$-thresholding is initialized by the solution of the soft-thresholding.
The regularization parameter $\lambda$ is selected by providing the best performance.

Fig. 4 shows the peak-signal noise ratio (PSNR) of recovery of
$\ell_q$-thresholding for different value of $q$ with SNR = 50 dB,
including the hard- and soft-thresholding as special cases.
It can be seen that each method is able to achieve a high PSNR greater than 60 dB
in recovering the synthetic image, but degrades significantly in recovering the
MRI image (less than 30 dB). This is due the nature that, as shown in Fig. 5,
the wavelet coefficients of the synthetic image are truly sparse (approximately 5.7\% nonzeros),
while that of a real-life image are not strictly sparse but rather approximately
follow an exponential decay, which is referred to as compressible.
Also due to this, for the two images, the best performance of $\ell_q$-thresholding are
given by $q=0.1$ and $q=0.7$, respectively, as shown in Fig. 4.
That is, a relatively small value of $q$ should be used for strictly sparse signals,
while a relatively large value of $q$ should be used for compressible (non-strictly sparse) signals.
Moreover, the hard- and $\ell_q$-thresholding can achieve significant
performance improvement over the soft-thresholding only for strictly sparse signals.

\subsection{Sparse Regression and Variable Selection}

Nowadays, the analysis of data sets with the number of variables comparable to or
even much larger than the sample size arises in many areas, such as genomics,
health sciences, economics and machine learning [126].
In this context with high-dimensional data, most traditional variable selection procedures,
such as AIC and BIC, becomes infeasible and impractical due to too expensive computational cost.
In this scenario, sparse regression, which can simultaneously select variables and estimate
coefficients of variables, has become a very popular topic in the last decade due to its
effectiveness in the high-dimensional case [37], [82].

Let ${\mathbf{X}} \in {\mathbb{R}^{m \times n}}$ denote the (deterministic) design matrix,
${\boldsymbol{\beta}} \in {\mathbb{R}^n}$ contains the unknown regression coefficients,
and ${\mathbf{y}} \in {\mathbb{R}^m}$ is the response vector.
Further, assume that ${\mathbf{y}}$ depends on ${\boldsymbol{\beta}}$ through a linear
combination ${\mathbf{X}}{\boldsymbol{\beta}}$ and the conditional log-likelihood given ${\mathbf{X}}$ is $\mathcal{L}(\boldsymbol{\beta} )$.
In the variable selection problem, the assumption is that majority of the
true regression coefficients are zero, and the goal is to identify and estimate the subset model.
Under the sparse assumption of the true regression coefficients,
a natural method for simultaneously locating and estimating those nonzero coefficients
in ${\boldsymbol{\beta}}$ is to maximize the following penalized likelihood of the form
\begin{equation}  
\mathop {{\mathrm{max}}}\limits_{\boldsymbol{\beta}}  \mathcal{L}({\boldsymbol{\beta}} ) - {P_\lambda }({\boldsymbol{\beta}} )
\end{equation}
where ${P_\lambda}$ is a generalized sparsity promotion penalty as defined in section 2.
Moreover, there exists a popular alternative which uses the following penalized least-square (LS) formulation
\begin{equation}  
\mathop {{\mathrm{min}}}\limits_{\boldsymbol{\beta}}  \frac{1}{2}\left\| {{\mathbf{y}} - {\mathbf{X}}{\boldsymbol{\beta}} } \right\|_{\mathrm{2}}^{\mathrm{2}} + {P_\lambda }({\boldsymbol{\beta}} ).
\end{equation}

\begin{figure*}[!t]
 \centering
 \includegraphics[scale = 0.3]{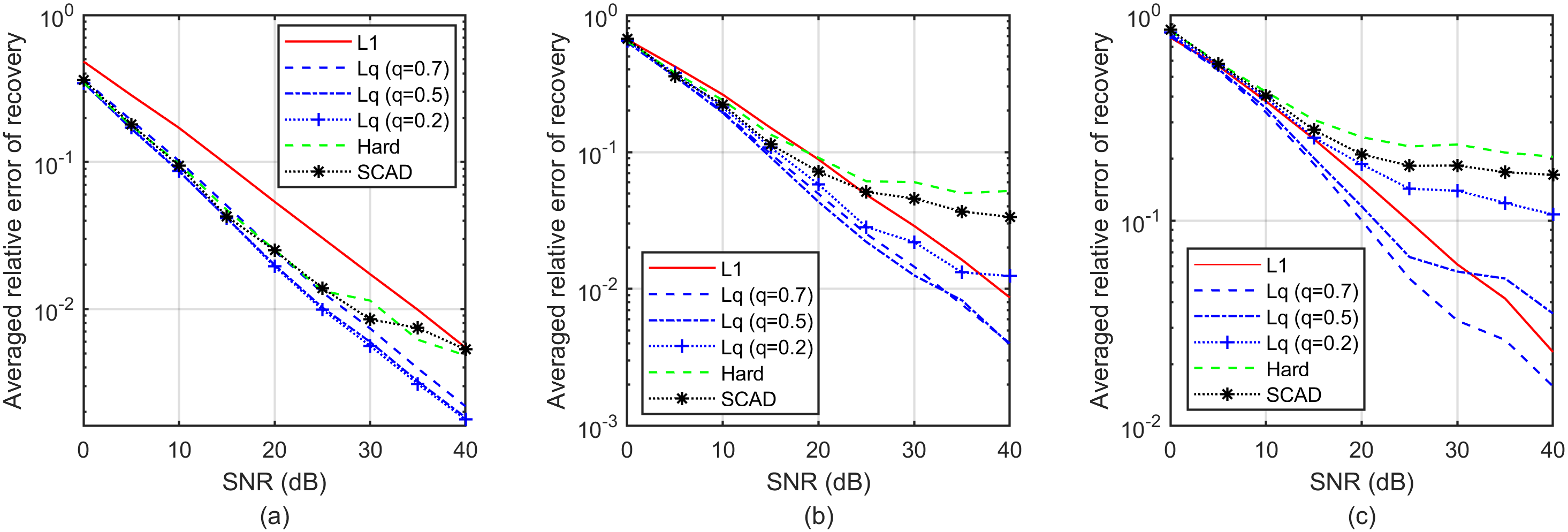}
\caption{Recovery performance of the hard-, SCAD-, $\ell_q$- and soft-thresholding versus SNR with zero initialization.
(a) 2\% active coefficients in $\boldsymbol{\beta}$, (b) 5\% active coefficients in $\boldsymbol{\beta}$, (c) 10\% active coefficients in $\boldsymbol{\beta}$.}
 \label{figure5}
\end{figure*}

\begin{figure*}[!t]
 \centering
 \includegraphics[scale = 0.3]{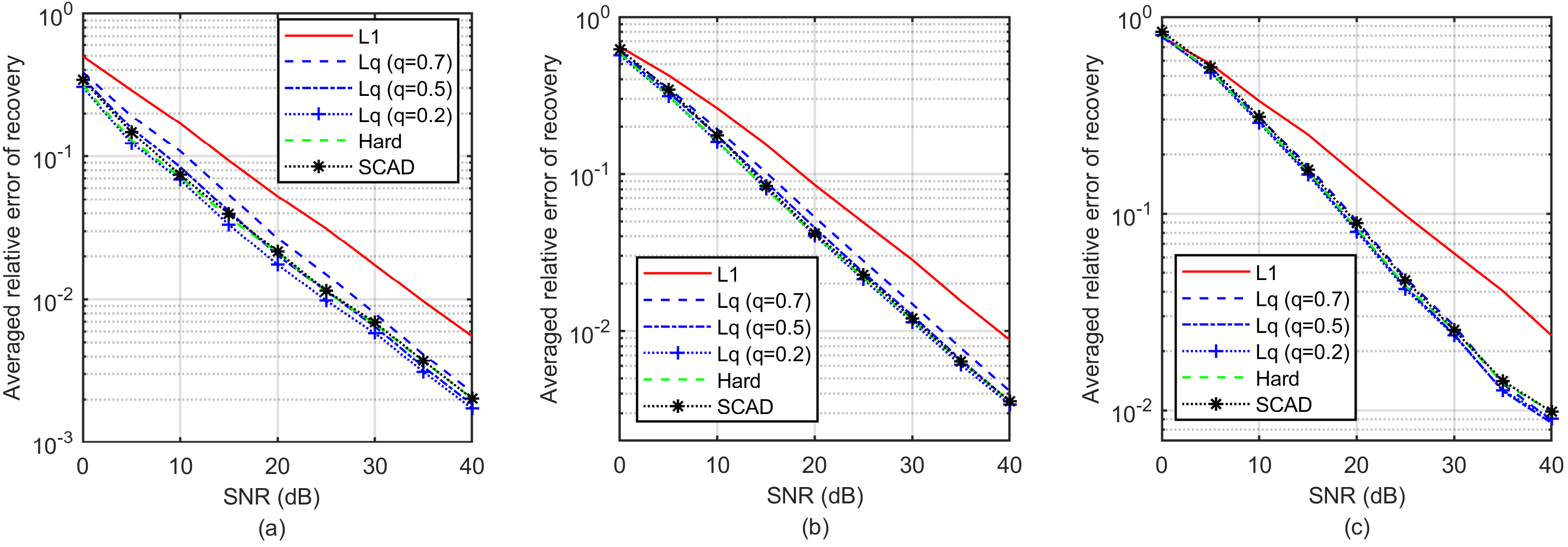}
\caption{Recovery performance of the hard-, SCAD-, $\ell_q$- and soft-thresholding versus SNR, initialized by the solution of the convex $\ell_1$ penalty. (a) 2\% active coefficients in $\boldsymbol{\beta}$, (b) 5\% active coefficients in $\boldsymbol{\beta}$, (c) 10\% active coefficients in $\boldsymbol{\beta}$.}
 \label{figure6}
\end{figure*}

The well-known LASSO method is first proposed in [37] for the linear regression problem (18).
In the same spirit of LASSO, nonconcave penalty functions, such as SCAD [11] and MCP [12],
have been proposed to select significant variables for various parametric models,
including linear regression, generalized linear regression and robust linear regression models [83].
Extension to some semiparametric models, such as the Cox model and partially linear models
have been considered in [84]--[86]. It has been shown in these works that,
with appropriately selected regularization parameters, nonconvex penalized estimators can
perform as well as the oracle procedure in selecting the correct subset model and estimating
the true nonzero coefficients. Further, even for super-polynomial of sample size,
nonconvex penalized likelihood methods possess model selection consistency with oracle properties [87].
In addition, adaptive LASSO has been proposed in [88], which uses an adaptively weighted $\ell_1$ penalty.
While LASSO variable selection can be inconsistent in some scenarios,
adaptive LASSO enjoys the oracle properties and also leads to a near-minimax optimal estimator.
The oracle property of adaptive LASSO has also been demonstrated in the high-dimensional case [89].

To solve a nonconvex penalized sparse regression problem,
locally approximation of the penalty function can be utilized,
such as local quadratic approximation (LQA) [11], [90] and local linear approximation (LLA) [91].
While the LQA based methods uses a backward stepwise variable selection procedure,
the LLA based method [91] is one-step and hence more favorable.

More recently, benefit from the progress in nonconvex and nonsmooth optimization,
direct methods have been widely designed. Specifically, in [92],
a coordinate-descent algorithm has been proposed for the linear regression model (18).
By using a continuation process of the parameters of the SCAD or MCP penalty,
its convergence to a minimum is guaranteed under certain conditions.
Meanwhile, an alternative coordinate-descent algorithm has been presented in [93]
with guaranteed convergence to a local minimum. Then, a cyclic descent algorithm employing
the $\ell_0$ penalty for multivariate linear regression has been introduced in [94].
Subsequently, a cyclic descent algorithm for the $\ell_q$ penalized LS problem has
been proposed in [7], [95], whilst an majorization-minimization (MM) algorithm with
momentum acceleration for the $\ell_0$ penalized LS formulation has been developed in [96].
For both the methods in [95] and [96], convergence to a local minimizer is guaranteed
under certain conditions. Moreover, as introduced in the last subsection,
there exist numerous algorithms, e.g., the PGD and ADMM algorithms,
can be applied to the penalized LS formulation (18).

In addition, a global optimization approach has been recently proposed in [206]
for concave penalties (e.g., SCAD and MCP) based nonconvex learning
via reformulating the nonconvex problems as general quadratic programs.
Meanwhile, theoretical analysis on the statistical performance of
local minimizer methods using folded concave penalties has been provided in [207].
Moreover, a class of nonconvex penalties termed \textit{trimmed Lasso} has been considered in [208],
which enables exact control of the desired sparsity level.

\textit{Example 2 (Estimation accuracy in different SNR and different sparsity level)}.
We evaluate the performance of different penalties for the linear sparse regression
problem (18) using the PGD (15), with $\boldsymbol{\beta} \in {\mathbb{R}^{256}}$ and
${\mathbf{X}} \in {\mathbb{R}^{100 \times 256}}$.
Three sparsity levels, 2\%, 5\%, and 10\% active coefficients of $\boldsymbol{\beta}$,
are considered. Fig. 6 shows the performance of different penalties versus SNR
when the PGD is initialized by zero, while Fig. 7 shows the results when the PGD
is initialized by the solution of the convex $\ell_1$ penalty.
It can be seen that the performance of the nonconvex penalties is heavily
dependent on the initialization. The advantage of a nonconvex penalty over the $\ell_1$ penalty
is significant when the sparsity level of the coefficient vector is relatively low and/or the SNR is relatively high.

\subsection{Sparse Signals Separation and Image Inpainting}

Sparse signals separation has wide applications, such as source separation,
super-resolution and inpainting, interference cancellation,
saturation and clipping restoration, and robust sparse recovery in impulsive (sparse) noise.
The objective of the problem is to demix the two sparse vectors ${{\mathbf{x}}_k} \in {\mathbb{R}^{{n_k}}}$, $k = 1,2$,
from their mixed linear measurements ${\mathbf{y}} \in {\mathbb{R}^m}$ as
\begin{equation}   
{\mathbf{y}} = {{\mathbf{A}}_{\mathrm{1}}}{{\mathbf{x}}_{\mathrm{1}}}{\mathrm{ + }}{{\mathbf{A}}_{\mathrm{2}}}{{\mathbf{x}}_{\mathrm{2}}}
\end{equation}
where ${{\mathbf{A}}_k} \in {\mathbb{R}^{m \times {n_k}}}$ are known deterministic dictionaries.
More specifics on the applications involving the model (19) are as follows.

\textit{1) Source separation}:
such as the separation of texture in images [97], [98]
and the separation of neuronal calcium transients in calcium imaging [99],
${{\mathbf{A}}_1}$ and ${{\mathbf{A}}_2}$ are two dictionaries allowing for
sparse representation of the two distinct features,
${{\mathbf{x}}_{\mathrm{1}}}$ and ${{\mathbf{x}}_{\mathrm{2}}}$ are the
(sparse or approximately sparse) coefficients describing these features [100]--[102].
\textit{2) Super-resolution and inpainting}:
in the super-resolution and inpainting problem for images, audio,
and video signals [103]--[105], only a subset of the desired signal
${{\mathbf{y}}_0} = {{\mathbf{A}}_{\mathrm{1}}}{{\mathbf{x}}_{\mathrm{1}}}$ is available.
Given ${\mathbf{y}}$, the objective is to fill in the missing parts in ${{\mathbf{y}}_0}$,
in which case ${{\mathbf{A}}_2} = {{\mathbf{I}}_m}$ and ${{\mathbf{x}}_{\mathrm{2}}}$ stands for the missing parts.
\textit{3) Interference cancellation}:
in some audio, video, or communication applications,
it is desired to restore a signal corrupted by narrowband interference,
such as electric hum [101]. As narrowband interference can be sparsely represented in the frequency domain,
${{\mathbf{A}}_2}$ can be an inverse discrete Fourier transform matrix.
\textit{4) Saturation and clipping restoration}:
in many practical systems where the measurements are quantized,
nonlinearities in amplifiers may result in signal saturation and
causes significant nonlinearity and potentially unbounded errors [101], [106], [107].
In this case, ${{\mathbf{y}}_0} = {{\mathbf{A}}_{\mathrm{1}}}{{\mathbf{x}}_{\mathrm{1}}}$ is the desired signal,
${\mathbf{y}}$ is the situated measurement with ${\mathbf{x}}_2$ stands for the saturation errors.
\textit{5) Robust sparse recovery in impulsive noise}:
impulsive noise is usually approximately sparse and has a distribution with heavy tail.
In practical image and video processing applications [108]--[110],
impulsive noise may come from the measurement process, or caused by transmission problems,
faulty memory locations, buffer overflow and unreliable memory [111]--[114].
In these cases, ${{\mathbf{x}}_{\mathrm{2}}}$ represents the (sparsely) impulsive noise and ${{\mathbf{A}}_2} = {{\mathbf{I}}_m}$.

Exploiting the sparsity, ${{\mathbf{x}}_{\mathrm{1}}}$ and ${{\mathbf{x}}_{\mathrm{2}}}$
can be reconstructed via the following formulation
\begin{equation}  
\begin{split}
&\mathop {{\mathrm{min}}}\limits_{{{\mathbf{x}}_1},{{\mathbf{x}}_2}} \mu {g_1}({{\mathbf{x}}_1}) + {g_2}({{\mathbf{x}}_2})\\
\mathrm{subject~ to~}&~~ {{\mathbf{A}}_1}{{\mathbf{x}}_1} + {{\mathbf{A}}_2}{{\mathbf{x}}_2} = {\mathbf{y}}~~~~~~~~~~~~~~~~~
\end{split}
\end{equation}
where ${g_1}$ and ${g_2}$ are penalties for sparsity promotion,
$\mu > 0$ is a parameter takes the statistic difference between the two components into consideration.
As will be shown later, when ${{\mathbf{x}}_{\mathrm{1}}}$ and ${{\mathbf{x}}_{\mathrm{2}}}$ have different sparsity properties,
using two different penalties for ${{\mathbf{x}}_{\mathrm{1}}}$ and ${{\mathbf{x}}_{\mathrm{2}}}$
can obtain performance gain over using a same one.

When both ${g_1}$ and ${g_2}$ are the ${\ell_1}$ penalty, i.e., ${g_1} = {g_2} = {\left\| \cdot \right\|_1}$,
(20) reduces to the sparse separation formulation in [102].
Further, when ${g_1} = {g_2} = {\left\| \cdot \right\|_1}$ and $\mu = 1$,
the formulation (20) degenerates to the BP form considered in [100].
Moreover, when ${{\mathbf{A}}_2} = {{\mathbf{I}}_m}$ and ${g_1} = {g_2} = {\left\|\cdot  \right\|_1}$,
(20) reduces to the ${\ell _1}$-regularized least-absolute problem for robust sparse recovery [115],
which has outstanding robustness in the presence of impulsive noise.
In addition, the M-estimation method in [218] can also be considered
as a special case of (20) with a selected robust loss term.

Naturally, nonconvex penalties can be expected to yield better reconstruction
performance over the above convex methods. For example, the ${\ell _{\mathrm{0}}}$ penalty
has been used in [116]--[119] to obtain more robust restoration of images corrupted by
salt-and-pepper impulsive noise. Very recently, a generalized formulation using ${\ell_q}$
penalty has been proposed in [120] with ${g_1} = \left\| \cdot \right\|_{{q_1}}^{{q_1}}$
and ${g_2} = \left\| \cdot \right\|_{{q_2}}^{{q_2}}$, $0 \le {q_{\mathrm{1}}},{q_2} < 1$.
The formulation (20) can be directly solved by a standard two-block ADMM procedure [29],
but it often fails to converge in the nonconvex case [120].
To develop convergent algorithms, a recenty work [120] proposed to solve a quadratic
approximation of (20) and developed two first-order algorithms based on the proximal
block coordinate descent (BCD) and ADMM frameworks.
The proximal BCD method consists of the following two update steps
\begin{align}
{\mathbf{x}}_1^{k + 1} &= {\mathrm{pro}}{{\mathrm{x}}_{(\beta \mu /{\eta _1}){g_1}}}\left\{ {{\mathbf{x}}_1^k - \frac{2}{{{\eta _1}}}{\mathbf{A}}_1^T({{\mathbf{A}}_1}{\mathbf{x}}_1^k + {{\mathbf{A}}_2}{\mathbf{x}}_2^k - {\mathbf{y}})} \right\}\\
{\mathbf{x}}_2^{k + 1} &= {\mathrm{pro}}{{\mathrm{x}}_{(\beta /{\eta _2}){g_2}}}\left\{ {{\mathbf{x}}_2^k - \frac{2}{{{\eta _2}}}{\mathbf{A}}_2^T({{\mathbf{A}}_1}{\mathbf{x}}_1^{k + 1} + {{\mathbf{A}}_2}{\mathbf{x}}_2^k - {\mathbf{y}})} \right\}
\end{align}
where ${\eta _1} > 0$ and ${\eta _2} > 0$ are proximal parameters, $\beta > 0$ is quadratic approximation parameter.
Suppose that ${g_1}$ and ${g_2}$ are closed, proper, lower semi-continuous, KL functions,
if ${\eta _1} > 2{\mathrm{ei}}{{\mathrm{g}}_{\max }}({\mathbf{A}}_1^T{{\mathbf{A}}_1})$ and
${\eta _2} > 2{\mathrm{ei}}{{\mathrm{g}}_{\max }}({\mathbf{A}}_2^T{{\mathbf{A}}_2})$,
the algorithm updated via (21) and (22) is a descent algorithm and the generated sequence
$\{({\mathbf{x}}_1^k,{\mathbf{x}}_2^k)\}$ converges to a stationary point of the approximated problem.

In many applications such as super-resolution and inpainting for color images with 3 (RGB) channels,
multichannel joint recovery is more favorable than independently recovery of each channel,
as the former can exploit the feature correlation among different channels.
In the multitask case, the linear measurements ${\mathbf{Y}} \in {\mathbb{R}^{m \times L}}$
of $L$ channels can be expressed as
\begin{equation}
{\mathbf{Y}} = {{\mathbf{A}}_{\mathrm{1}}}{{\mathbf{X}}_{\mathrm{1}}}{\mathrm{ + }}{{\mathbf{A}}_{\mathrm{2}}}{{\mathbf{X}}_{\mathrm{2}}}
\end{equation}
where ${{\mathbf{X}}_k} \in {\mathbb{R}^{{n_k} \times L}}$, $k = 1,2$, are the sparse features in the two components.
To exploit the joint sparsity among the $L$ channels, joint sparsity penalties can be used, which is defined as
\begin{equation}
\begin{split}
{\tilde P_\lambda }({\mathbf{X}}) &= \sum\nolimits_i {{P_\lambda }\left( {{{\left\| {{\mathbf{X}}[i,:]} \right\|}_2}} \right)}  \\
&= \sum\nolimits_i {{P_\lambda }\left( {{{\left( {\sum\nolimits_j {{{\mathbf{X}}^2}[i,j]} } \right)}^{1/2}}} \right)}    \notag .
\end{split}
\end{equation}

\begin{figure*}[!t]
\centering
\subfloat[Inpainting of a $318\times 500$ image corrupted by overwritten text]{
{\includegraphics[scale = 0.24]{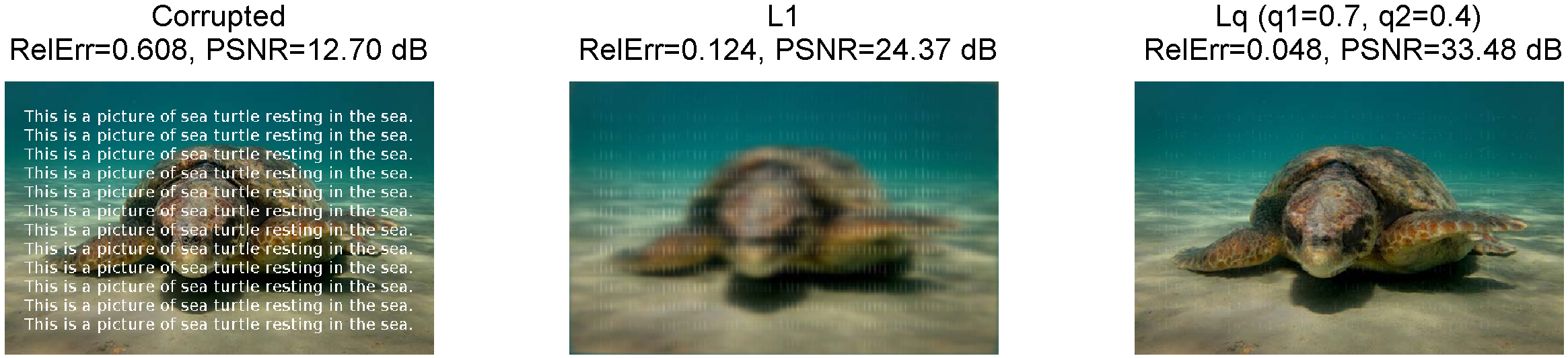}}}\\
\subfloat[Inpainting of a $512\times 512$ image corrupted by salt-and-pepper noise (30\% of the pixels are corrupted)]{
{\includegraphics[scale = 0.26]{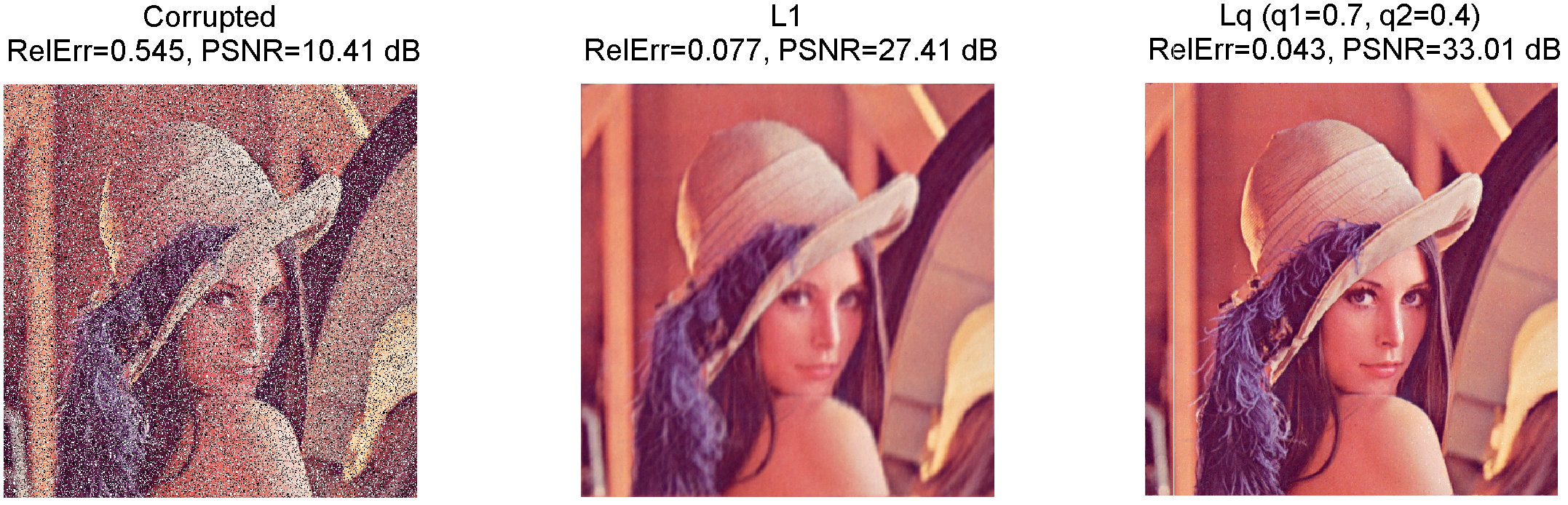}}}
\caption{Restoration performance of the $\ell_1$ and $\ell_q$ regularization in color images inpainting.}
\label{figure7}
\end{figure*}

\begin{figure*}[!t]
 \centering
 \includegraphics[scale = 0.19]{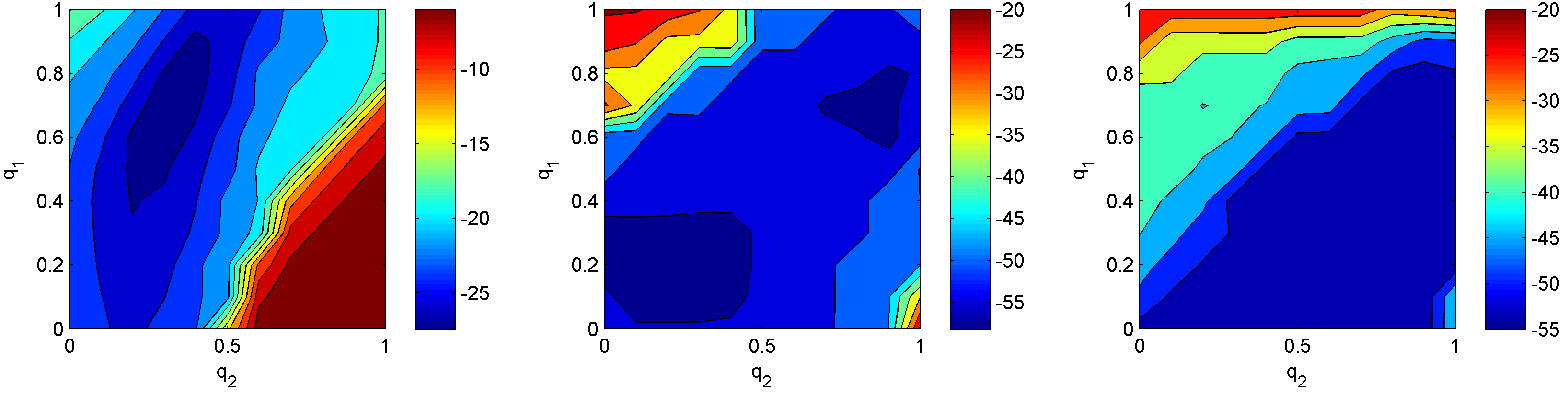}
\caption{Restoration performance versus $q_1$ and $q_2$ in terms of RelErr in dB defined as $20{\log _{10}}(\|{{\mathbf{\hat x}}_1} - {{\mathbf{x}}_1}\|{_2}/\|{{\mathbf{x}}_1}\|{_2})$. \textit{Left}: Case 1: restoration of the image in Fig. 8 (a), ${{\mathbf{X}}_{\mathrm{1}}}$ contains the DCT coefficients of the image and ${{\mathbf{X}}_{\mathrm{2}}}$ represents the overwritten text. \textit{Middle}: Case 2: both ${{\mathbf{x}}_{\mathrm{1}}}$ and ${{\mathbf{x}}_{\mathrm{2}}}$ are strictly sparse. \textit{Right}: Case 3: ${{\mathbf{x}}_{\mathrm{1}}}$ is strictly sparse and ${{\mathbf{x}}_{\mathrm{2}}}$ is $\alpha$-stable (${\mathrm{S}}\alpha {\mathrm{S}}$) noise (non-strictly sparse).}
 \label{figure8}
\end{figure*}

Using such a penalty, e.g., ${G_1}$ for ${{\mathbf{X}}_{\mathrm{1}}}$ and ${G_2}$ for ${{\mathbf{X}}_{\mathrm{2}}}$,
the proxiaml BCD algorithm for the multitask problem consists of the following two steps [120]
\begin{equation}
\begin{split}
&{\mathbf{X}}_1^{k + 1}= \arg \mathop {\min }\limits_{{{\mathbf{X}}_1}}\mu {G_1}({{\mathbf{X}}_1}) + \\
&~~~ \frac{{{\eta _3}}}{{2\beta }}\left\| {{{\mathbf{X}}_1} - {\mathbf{X}}_1^k + \frac{2}{{{\eta _3}}}{\mathbf{A}}_1^T({{\mathbf{A}}_1}{\mathbf{X}}_1^k + {{\mathbf{A}}_2}{\mathbf{X}}_2^k - {\mathbf{Y}})} \right\|_\mathrm{F}^2\\
\end{split}
\end{equation}
\begin{equation}
\begin{split}
&{\mathbf{X}}_2^{k + 1}= \arg \mathop {\min }\limits_{{{\mathbf{X}}_2}} {G_2}({{\mathbf{X}}_2}) + \\
&~~ \frac{{{\eta _4}}}{{2\beta }}\left\| {{{\mathbf{X}}_2} - {\mathbf{X}}_2^k + \frac{2}{{{\eta _4}}}{\mathbf{A}}_2^T({{\mathbf{A}}_1}{\mathbf{X}}_1^{k + 1} + {{\mathbf{A}}_2}{\mathbf{X}}_2^k - {\mathbf{Y}})} \right\|_\mathrm{F}^2
\end{split}
\end{equation}
where ${\eta_3} > 0$ and ${\eta _4} > 0$ are proximal parameters.
These two subproblems can be solved row-wise as (5).
Suppose that ${G_1}$ and ${G_2}$ are closed, proper, lower semi-continuous,
KL functions, if ${\eta _3} > 2{\mathrm{ei}}{{\mathrm{g}}_{\max }}({\mathbf{A}}_1^T{{\mathbf{A}}_1})$,
and ${\eta _4} > 2{\mathrm{ei}}{{\mathrm{g}}_{\max }}({\mathbf{A}}_2^T{{\mathbf{A}}_2})$,
the algorithm updated via (24) and (25) is a descent algorithm and the generated
sequence $\{ ({\mathbf{X}}_1^k,{\mathbf{X}}_2^k)\} $ converges to a stationary point
of the approximated problem.

\textit{Example 3 (Color image inpainting)}.
We compare the performance of the ${\ell _q}$ and ${\ell _1}$ regularization on
color images inpainting using the BCD method (24) and (25).
The task is to restore the original image from text overwriting or salt-and-pepper noise corruption.
In this case, ${{\mathbf{A}}_2} = {\mathbf{I}}$ and ${{\mathbf{X}}_{\mathrm{2}}}$ represents the
sparse corruption in the three (RGB) channels.
We select ${{\mathbf{A}}_1}$ as an inverse discrete cosine transformation (IDCT) matrix,
and, accordingly, ${{\mathbf{X}}_1}$ contains the DCT coefficients of the image.
As shown in Fig. 8, ${\ell _q}$ regularization significantly outperforms the ${\ell_1}$ one,
e.g., the improvement is more than 9 dB in the overwritten text case
and more than 5 dB in the salt-and-pepper noise case.

\textit{Example 4 (Evaluation on different penalties)}.
An important problem in practice is how to select a nonconvex penalty.
This example sheds some light on this by three application cases.
Case 1: the color image inpainting experiment in Fig. 8 (a).
Case 2: $L=1$, ${{\mathbf{A}}_1} \in {\mathbb{R}^{128 \times 128}}$ and
${{\mathbf{A}}_2} \in {\mathbb{R}^{128 \times 128}}$ are respectively DCT and Gaussian matrices,
${{\mathbf{x}}_{\mathrm{1}}}$ and ${{\mathbf{x}}_{\mathrm{2}}}$ are strictly sparse vectors with sparsity $K = 25$.
Case 3 (robust sparse recovery in impulsive noise):
$L = 1$, ${{\mathbf{A}}_1} \in {\mathbb{R}^{100 \times 256}}$ is a Gaussian matrix,
${{\mathbf{A}}_2} = {{\mathbf{I}}_{100}}$, ${{\mathbf{x}}_{\mathrm{1}}}$ is a strictly sparse vector with $K = 20$,
and ${{\mathbf{x}}_{\mathrm{2}}}$ is symmetric $\alpha$-stable (${\mathrm{S}}\alpha {\mathrm{S}}$)
noise with characteristic exponent $\alpha = 1$ and dispersion $\gamma = {10^{ - 3}}$.

As the ${\ell_q}$ penalty has a flexible parametric form that adapts to
different thresholding functions, we evaluate the effect of the values of
$q_1$ and $q_2$ on the recovery performance in the three cases.
Fig. 9 indicates that for strictly sparse signals, a relatively small value
of $q$ is favorable, while for non-strictly sparse signals, a relatively large value
of $q$ is favorable. For example, in case 1,
${{\mathbf{X}}_{\mathrm{1}}}$ (DCT coefficients of the image) is non-strictly sparse
whilst ${{\mathbf{X}}_2}$ is strictly sparse, thus, a relatively large value of
$q_1$ and a relatively small value of $q_2$ should be used.
In case 3, ${{\mathbf{x}}_{\mathrm{1}}}$ is strictly sparse whilst
${{\mathbf{x}}_{\mathrm{2}}}$ (${\mathrm{S}}\alpha {\mathrm{S}}$ noise) is non-strictly sparse,
thus, a relatively small value of $q_1$ and a relatively large value of $q_2$
would result in good performance.

\subsection{Sparse PCA}

PCA is a useful tool for dimensionality reduction and feature extraction,
which has been applied in virtually all areas of science and engineering,
such as signal processing, machine learning, statistics, biology, medicine,
finance, neurocomputing, and computer networks, to name just a few.
In many real applications, sparse loading vectors are desired in PCA to
enhance the interpretability of the principle components (PCs).
For instance, in gene analysis, the sparsity of PCs can facilitate the
understanding of the relation between the whole gene microarrays and certain genes;
in financial analysis, the sparsity of PCs implies fewer assets in a portfolio
thus is helpful to reducing the trading costs. In these scenarios,
it is desirable not only to achieve the dimensionality reduction
but also to reduce the number of explicitly used variables.

To achieve sparse PCA, as hoc methods were firstly designed via thresholding
the PC loadings [124], [125]. Then, more popular methods incorporating a sparsity
inducing mechanism into the traditional PCA formulation have been developed [126].
Let ${\mathbf{X}} \in {\mathbb{R}^{d \times n}}$ be a centered data matrix with
$d$ and $n$ respectively be the dimensionality and the size of the data.
Employing a sparsity constraint in the traditional PCA model,
the sparse PCA problem to find an $m$-dimensional subspace
can be formulated as
\begin{equation} 
\begin{split}
&{\mathbf{w}}_i^* = \arg \mathop {{\mathrm{max}}}\limits_{{{\mathbf{w}}_i} \in {\mathbb{R}^d}} \left\| {{\mathbf{w}}_i^T{\mathbf{X}}} \right\|_2^2\\
\mathrm{subject~ to~} &~{\mathbf{w}}_i^T{{\mathbf{w}}_j} = {\delta _{i,j}}, P({{\mathbf{w}}_i}) \le k ~~~~~~~~~~~~
\end{split}
\end{equation}
for $i,j = 1, \cdots ,m$. $P$ is a sparsity inducing penalty.
Instead of solving (26), a widely used greedy strategy is to find
the approximate solution of (26) by sequentially solving the following single PC problem
\begin{equation} 
\begin{split}
&{\mathbf{w}}_{}^* = \arg \mathop {{\mathrm{max}}}\limits_{{\mathbf{w}} \in {^d}} \left\| {{\mathbf{w}}_{}^T{\mathbf{X}}} \right\|_2^2\\
\mathrm{subject~ to~}&~ {\left\| {\mathbf{w}} \right\|_2} \le 1, P({\mathbf{w}}) \le k.~~~~~~~~~~~~~~~~~~
\end{split}
\end{equation}
An alternative of the sparsity constrained formulation (27) is the sparsity penalized formulation as follows
\begin{equation} 
\begin{split}
&{\mathbf{w}}_{}^* = \arg \mathop {{\mathrm{max}}}\limits_{{\mathbf{w}} \in {^d}} \left\| {{\mathbf{w}}_{}^T{\mathbf{X}}} \right\|_2^2 - {P_\lambda }({\mathbf{w}})\\
\mathrm{subject~ to~}&~ {\left\| {\mathbf{w}} \right\|_2} \le 1 ~~~~~~~~~~~~~~~~
\end{split}
\end{equation}
where $\lambda>0$ is a regularization parameter.

There exist numerous algorithms employing different penalty functions for the
formulations (26)--(28) and their variants. In [126], the ${\ell _1}$ constraint
is used for the formulation (26). In [127], an elastic-net regression based algorithm has been proposed.
Then, a semidefinite relaxation method has been developed in [128], [129] for the formulation (27) with ${\ell _0}$ constraint.
Meanwhile, in [130], a regularized low rank matrix approximation method has been designed
with the consideration of different penalties, i.e., ${\ell _1}$, ${\ell _0}$ and SCAD.
In [131], with ${\ell_1}$ and ${\ell_0}$ penalties, reformulations of (28) and its block variant
have been solved via gradient algorithms.
Generalized version of (28) with ${\ell_0}$ penalty has been considered in [132].
Moreover, an alternative discrete spectral formulation of ${\ell_0}$ constrained (27)
and an effective greedy approach have been presented in [133].
In [134], unified alternating maximization framework for ${\ell_0}$ and ${\ell_1}$
constrained or penalized sparse PCA problems (using ${\ell_1}$ or ${\ell_2}$ loss) has been proposed.

More recently, robust sparse PCA using ${\ell_1}$ loss and ${\ell_q}$ penalty has been considered in [135].
Meanwhile, an ADMM based distributed sparse PCA algorithm has been proposed in [136]
which covers the ${\ell_1}$, log sum and MCP penalties.
Moreover, Shatten-$q$ penalty has been used for structured sparse PCA in [137].
In addition, there also exist several other methods for ${\ell_0}$ constrained
or penalized sparse PCA problems, e.g., [138]--[141].

\section{Sparse Matrix Recovery}

This section reviews the nonconvex regularization based sparse matrix recovery problems,
mainly on large covariance matrix and inverse covariance matrix estimation,
which are two fundamental problems in modern multivariate analysis [142].
Nowadays, the advance of information technology makes massive high-dimensional
data widely available for scientific discovery.
In this context, effective statistical analysis for high-dimensional data is becoming increasingly important.
In many applications involving statistical analysis of high-dimensional data,
estimating large covariance or inverse covariance matrices is necessary for
effective dimensionality reduction or discriminant analysis.
Such applications arise in economics and finance, bioinformatics, social networks,
smart grid, climate studies, and health sciences [142]--[144].
In the high-dimensional setting, the dimensionality is often comparable to
(or even larger than) the sample size. In these cases, the sample covariance matrix estimator
has a poor performance [145], and intrinsic structures such as sparsity can be
exploited to improve the estimation of covariance and inverse covariance matrices [142], [146]--[148], [155].

\subsection{Large Sparse Covariance Matrix Estimation}

Consider a vector ${\mathbf{x}} \in {\mathbb{R}^d}$ with covariance ${\boldsymbol{\Sigma}} = E\{ {\mathbf{x}}{{\mathbf{x}}^T}\}$,
the objective is to estimate its covariance from $n$ observations ${{\mathbf{x}}_1}, \cdots ,{{\mathbf{x}}_n}$.
Usually, compared to estimate ${\boldsymbol{\Sigma}}$ directly, it is more favorable to estimate the correlation matrix first,
${\mathbf{R}} = {\mathrm{diag}}{({\boldsymbol{\Sigma}} )^{ - 1/2}}{\boldsymbol{\Sigma}} {\mathrm{diag}}{({\boldsymbol{\Sigma}} )^{ - 1/2}}$.
Then, given the estimated correlation matrix ${\hat{\mathbf{ R}}}$, the corresponding estimation of
the covariance matrix is $\hat {\boldsymbol{\Sigma}}  = {\mathrm{diag}}{({\mathbf{S}})^{1/2}}{\hat{\mathbf{ R}}}{\mathrm{diag}}{({\mathbf{S}})^{1/2}}$,
where ${\mathbf{S}}$ is the sample covariance matrix.
That is because the correlation matrix has the same sparsity pattern of the covariance matrix
but with all the diagonal elements known to be one, thus,
it can be estimated more accurately than the covariance matrix [149]--[151].

Given the sample correlation matrix ${\mathbf{S}}$, the generalized thresholding estimator [148] solves the following problem
\begin{equation}
\mathop {{\mathrm{min}}}\limits_{\mathbf{R}} \frac{1}{2}\left\| {{\mathbf{R}} - {\mathbf{S}}} \right\|_\mathrm{F}^2 + \sum\nolimits_{i \ne j} {{P_\lambda }({{\mathbf{R}}_{ij}})}
\end{equation}
with ${P_\lambda}$ be generalized penalty function for sparsity promotion as introduced in section 2.
Note that, the diagonal elements are not penalized since the diagonal elements of a correlation
(also covariance) matrix are always positive. The solution to (29) is a thresholding of the sample correlation matrix ${\mathbf{S}}$,
which can be efficiently computed as shown in section 2.

The thresholding estimator (29) have good theoretical properties.
It is consistent over a large class of (approximately) sparse covariance matrices [148].
However, in practical finite sample applications, such an estimator is not always positive-definite although
it converges to a positive-definite limit in the asymptotic setting [149], [151].
To simultaneously achieve sparsity and positive-definiteness, positive-definite constraint can be added into (29) as [152]
\begin{equation} 
\begin{split}
&\mathop {{\mathrm{min}}}\limits_{\mathbf{R}} \frac{1}{2}\left\| {{\mathbf{R}} - {\mathbf{S}}} \right\|_\mathrm{F}^2 + \sum\nolimits_{i \ne j} {{P_\lambda }({{\mathbf{R}}_{ij}})} ~~~~~~\\
\mathrm{subject~ to~~}&~ {\mathrm{diag}}({\mathbf{R}}) = {{\mathbf{I}}_d} ~\mathrm{and}~ {\mathbf{R}} \ge \varepsilon {{\mathbf{I}}_d}
\end{split}
\end{equation}
where $\varepsilon > 0$ is the lower bound for the minimum eigenvalue.
An alternating minimization algorithm has been proposed for (30) in [152],
which is guaranteed to be globally convergent for a generalized nonconvex penalty ${P_\lambda}$
(when it is a closed, proper, lower semi-continuous, KL function).

Suppose the ``approximately sparse'' covariance matrix satisfies
\begin{equation}
\begin{split}
&\mathcal{U}(\kappa ,p,{M_d},\xi )\\
&: =\left\{ {{\boldsymbol{\Sigma}} :\mathop {\max }\limits_i {{\boldsymbol{\Sigma}} _{ii}} \le \kappa ,{{\tilde {\boldsymbol{\Sigma}} }^{ - 1}}{\boldsymbol{\Sigma}} {{\tilde {\boldsymbol{\Sigma}} }^{ - 1}} \in { \mathcal{M}}(p,{M_d},\xi )} \right\}   \notag
\end{split}
\end{equation}
where $\tilde {\boldsymbol{\Sigma}}  = {\mathrm{diag}}(\sqrt {{{\boldsymbol{\Sigma}} _{11}}} , \cdots ,\sqrt {{{\boldsymbol{\Sigma}} _{dd}}} )$ and
\begin{equation}
\begin{split}
&{ \mathcal{M}}(p,{M_d},\xi )\\
&: = \left\{ {{\mathbf{R}}: \mathop {\max }\limits_i \sum\limits_{j \ne i} {{{\left| {{{\mathbf{R}}_{ij}}} \right|}^p} \le {M_d}} ,
{{{\mathbf{R}}_{ii}} = 1,{\mathrm{ei}}{{\mathrm{g}}_{\min }}({\mathbf{R}}) = \xi }} \right\}  \notag
\end{split}
\end{equation}
with $0\leq p<1$. It has been shown in [152] that, for a generalized nonconvex penalty
which satisfies (3) and for large enough $n$, the positive-definite estimator (30) satisfies
\begin{align}
{\left\| {\hat {\boldsymbol{\Sigma}} - {\boldsymbol{\Sigma}} } \right\|_2} = {O_P}\left( {{M_d}{{\left( {\frac{{\log d}}{n}} \right)}^{\frac{{1 - p}}{2}}}} \right) \notag
\end{align}
which achieves the minimax lower bound over the class $\mathcal{U}(\kappa ,p,{M_d},\xi )$ under the Gaussian model [154],
as the estimator (29). The estimators (29) and (30) give the same estimation with overwhelming probability in the asymptotic case.
Thus, for covariance matrices in $\mathcal{U}(\kappa ,p,{M_d},\xi )$, both the estimators (29) and (30)
are asymptotically rate optimal under the flexible elliptical model.

\begin{figure}[!t]
\centering
\subfloat[ ]{
{\includegraphics[scale = 0.2]{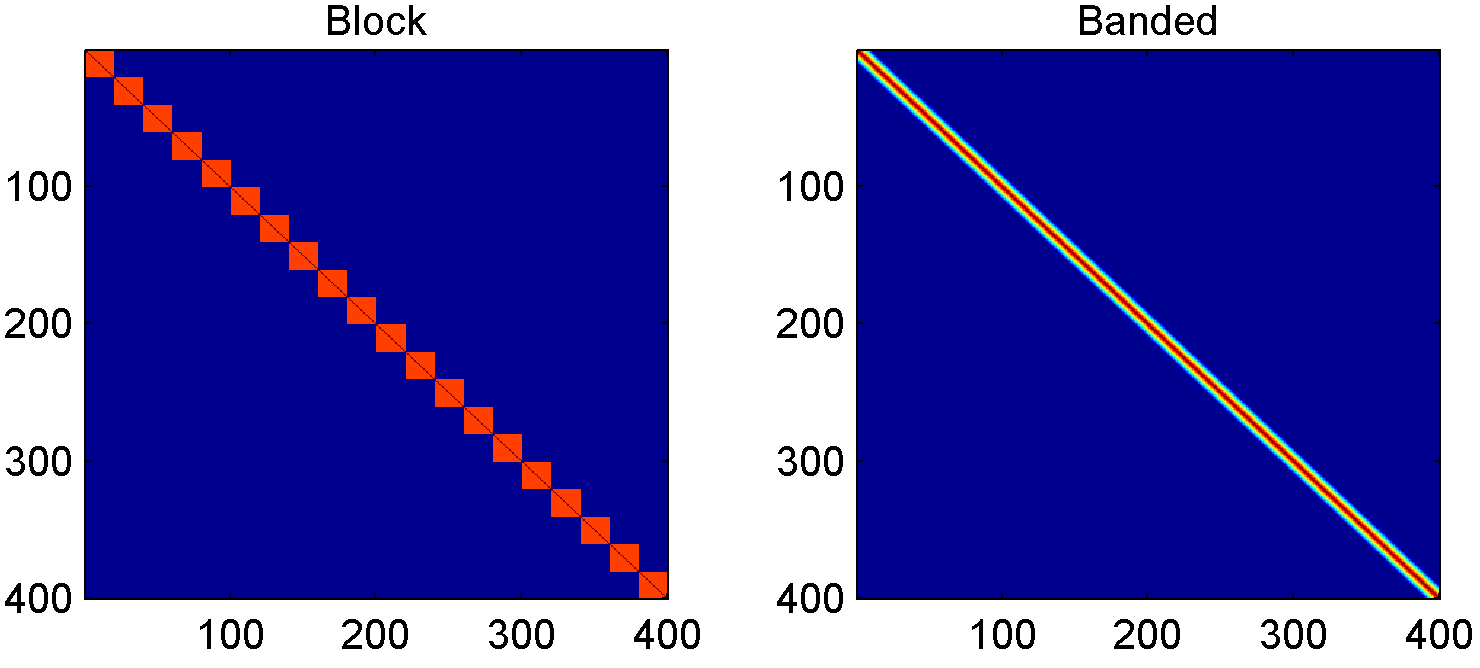}}}\\
\subfloat[ ]{
{\includegraphics[scale = 0.24]{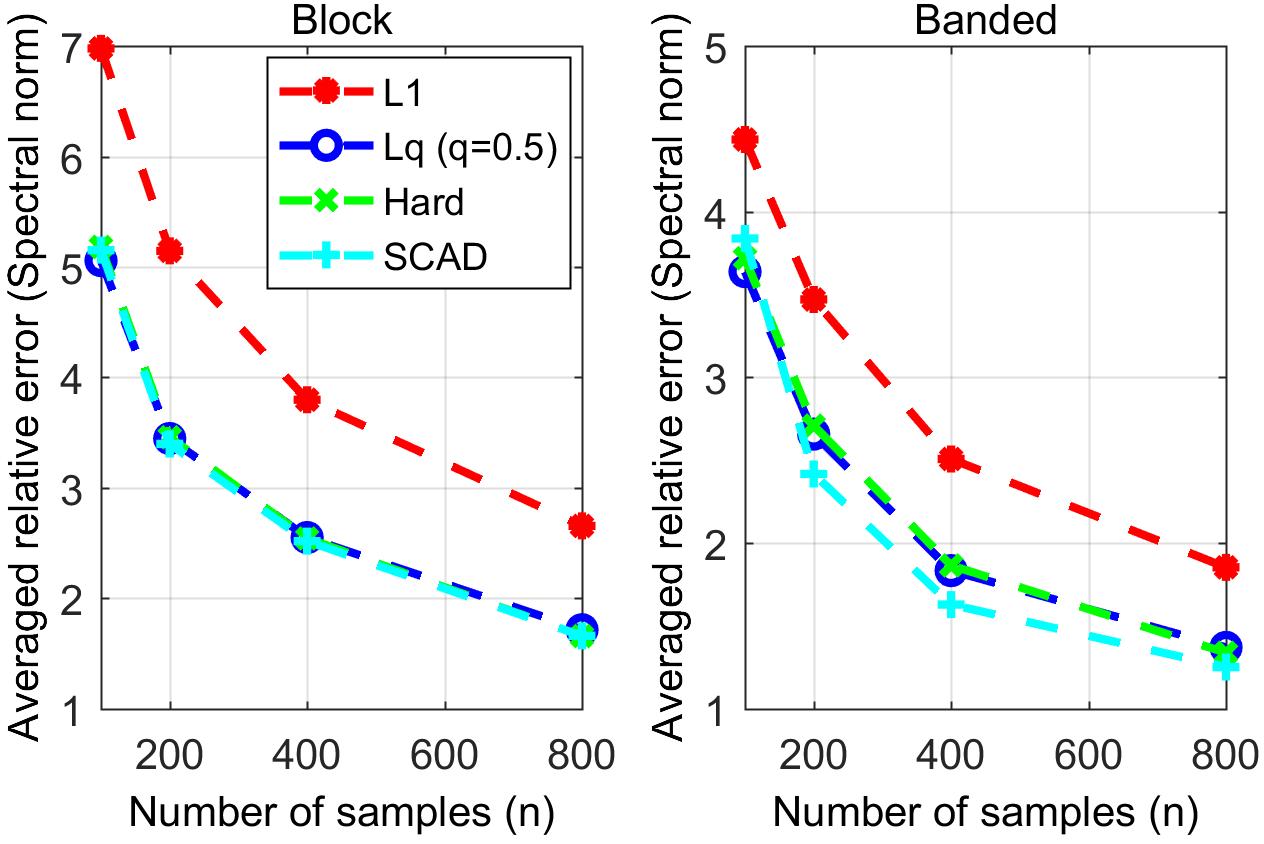}}}
\caption{(a) Heat maps of the two simulated covariance matrices. (b) Performance of different penalties in estimating the two simulated covariance matrices.}
\label{figure9}
\end{figure}

\begin{figure*}[!t]
 \centering
 \includegraphics[scale = 0.23]{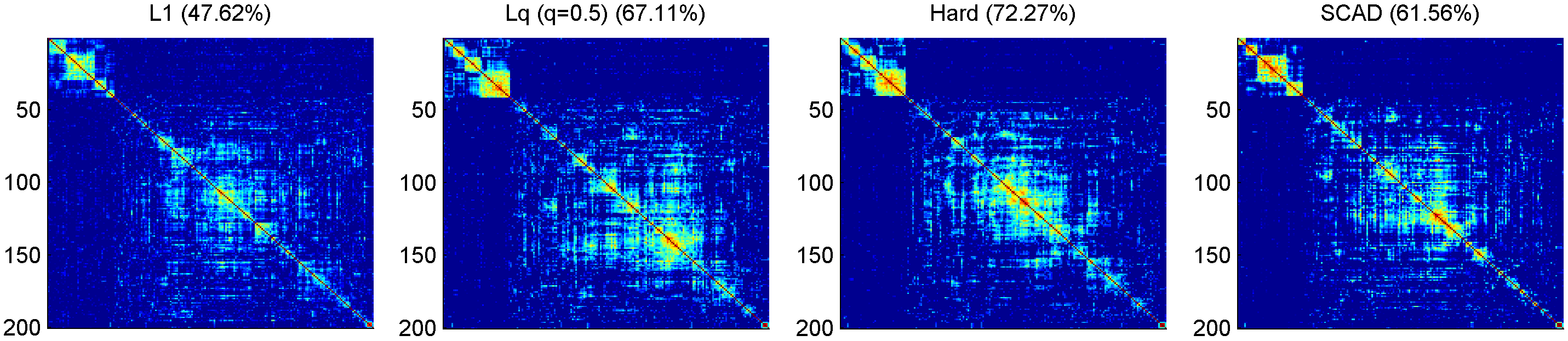}
\caption{Heat maps of the absolute values of estimated correlations for the selected 200 genes (the percentage of the entries with absolute values less than ${10^{-5}}$ is given in parentheses).}
 \label{figure10}
\end{figure*}

\textit{Example 5 (Estimation on simulated datasets)}.
We evaluate the alternating minimization algorithm [152] solving (30) $(\varepsilon={10^{ - 3}}) $
with different penalties in terms of relative error of estimation under spectral norm.
Each provided result is an average over 100 independent runs.
Two typical sparse covariance matrix models, block and banded of size $d = 400$, are considered.
Fig. 10 demonstrates that, the nonconvex SCAD, hard- and ${\ell_q}$-penalties can yield
considerable performance gain over the ${\ell_1}$ penalty.

\textit{Example 6 (Gene clustering)}.
We further consider a gene clustering example using a gene expression dataset from
a small round blue-cell tumors (SRBCTs) microarray experiment [153].
This dataset contains 88 SRBCT tissue samples, and 2308 gene expression values are recorded for each sample.
We use the 63 labeled calibration samples and pick up the top 40 and bottom 160 genes based on their F-statistic.
Accordingly, the top 40 genes are informative while the bottom 160 genes are non-informative.
Fig. 11 shows the heat maps of the absolute values of estimated correlations
by the compared penalties for the selected 200 genes.
Each heat map is ordered by group-average-agglomerative clustering based on the estimated correlation.
It can be seen that, compared with the ${\ell_1}$ penalty, each nonconvex penalty can give
cleaner and more informative estimates of the sparsity pattern.

\subsection{Large Sparse Inverse Covariance Matrix Estimation}

Large inverse covariance matrix estimation is another fundamental problem in modern multivariate analysis.
While the covariance matrix ${\boldsymbol{\Sigma}}  = E\{ {\mathbf{x}}{{\mathbf{x}}^T}\} \in {\mathbb{R}^{d \times d}}$
only captures the marginal correlations among the variables in $\mathbf{x}$,
the inverse covariance matrix $\boldsymbol{\Theta} = {{\boldsymbol{\Sigma}}^{-1}}$ captures the conditional correlations
among these variables and is closely related to undirected graphs under a Gaussian model.
Following the parsimony principle, it is desirable to choose the simplest model
(i.e., the sparsest graph) that adequately explains the data.
To achieve this, sparsity promotion can be used to improve the interpretability
of the model and prevent overfitting. In addition to the graphical model problem,
the interest in sparse inverse covariance estimation also arises in many other areas
such as high dimensional discriminant analysis, portfolio allocation,
and principal component analysis [142], [156]--[158].

One of the most popular approaches of estimating sparse inverse
covariance matrices is through the penalized maximum likelihood.
Specifically, assume that ${{\mathbf{x}}_1}, \cdots ,{{\mathbf{x}}_n}$
are independently and identically Gaussian distributed with zero-mean and covariance ${\boldsymbol{\Sigma}}$,
the negative log-likelihood function is $\mathcal{L}({\boldsymbol{\Theta}} ) = {\mathrm{tr}}({\mathbf{S}}{\boldsymbol{\Theta}} )- \log |{\boldsymbol{\Theta}} |$,
where $\mathbf{S}$ is the sample covariance. Then, the sparsity penalized likelihood estimator is given by
\begin{align}  
\mathop {{\mathrm{min}}}\limits_{\boldsymbol{\Theta}}  {\mathrm{tr}}({\mathbf{S}}{\boldsymbol{\Theta}} ) - \log |{\boldsymbol{\Theta}} | + \sum\nolimits_{i \ne j}^{} {{P_\lambda }({{\boldsymbol{\Theta}} _{ij}})} .
\end{align}
The positive-definiteness of this estimator is naturally ensured thanks to the logarithmic barrier term.

As the loss term (the log-likelihood $\mathcal{L}({\boldsymbol{\Theta}})$) in (31) is convex,
when ${P_\lambda}$ is the $\ell_1$ penalty, the formulation (31) is convex
and efficient ADMM algorithm can be applied with guaranteed convergence [29], [159].
However, with a nonconvex ${P_\lambda}$,
there is no guarantee of the convergence for such an algorithm,
since the gradient of the loss is not Lipschitz continuous.

For a folded concave penalty, such as the SCAD, $\ell_q$, or MCP penalty,
a strategy is to use local quadratic approximation (LQA) or
local linear approximation (LLA) of the penalty [160]--[162].
In such a manner, a concave penalized problem is converted
into a sequence of reweighed $\ell_1$ penalized problems.
Very recently, direct methods have been developed. Specifically,
a block cyclic descent (CD) algorithm has been proposed for the
formulation (31) with $\ell_q$ penalty in [163], [164]. Subsequently,
a coordinate-by-coordinate CD algorithm with guaranteed convergence
(to a local minimizer) has been proposed in [165] for
the $\ell_0$ penalized log-likelihood formulation
\begin{align}
\mathop {{\mathrm{min}}}\limits_{\boldsymbol{\Theta}}  {\mathrm{tr}}({\mathbf{S}}{\boldsymbol{\Theta}} ) - \log |{\boldsymbol{\Theta}} | + \lambda {\left\| {\boldsymbol{\Theta}}  \right\|_0}.
\end{align}
Meanwhile, a highly efficient block iterative method for (32) has been developed in [166],
which is capable of handling large-scale problems (e.g., $d = {10^4}$).
Moreover, extension of the $\ell_0$ penalized formulation (32) for
time-varying sparse inverse covariance estimation for the application
of tracking dynamic functional magnetic resonance imaging (fMRI)
brain networks has been considered in [167].

With a general nonconvex penalty function,
the rate of convergence for estimating sparse inverse covariance matrices
based on penalized likelihood has been established in [162],
which under the Frobenius norm is of order $\sqrt {s \cdot \log d/n}$,
where $s$ and $d$ are respectively the number of nonzero elements and size of the matrix,
and $n$ is the sample size. Further, while the $\ell_1$-penalty can achieve simultaneously
the optimal rate of convergence and sparsistency when the number of nonzero off-diagonal
entries is no larger than $O(d)$, there is no need of such a restriction for an unbiased nonconvex penalty,
e.g., the SCAD or hard-thresholding penalty.

\section{Low-Rank Matrix Recovery}

This section reviews nonconvex regularization based low-rank recovery problems,
mainly on matrix completion and robust PCA, which are two hot topics in the past few years.
Matrix completion aims to recover a low-rank matrix from partially observed entries,
while robust PCA aims to decompose a low-rank matrix from sparse corruption.

\subsection{Matrix Completion}
Matrix completion problems deal with the recovery of a matrix from
its partially observed (may be noisy) entries [168]--[171],
which has found applications in various fields such as recommender systems [172],
computer vision [173] and system identification [174], to name a few.
The goal of matrix completion is to recover a matrix ${\mathbf{X}} \in {\mathbb{R}^{m \times n}}$
from its partially known entries $\{ {{\mathbf{M}}_{i,j}={\mathbf{X}}_{i,j}}:(i,j) \in \Omega \} $,
where $\Omega  \subset [1, \cdots ,m] \times [1, \cdots ,n]$ is a random subset.
This can be achieved via exploiting the low-rankness of $\mathbf{X}$ by the following formulation
\begin{equation}
\begin{split}
&\mathop {{\mathrm{min}}}\limits_{\mathbf{X}} \bar P({\mathbf{X}})\\
\mathrm{subject~ to~}&~~{{ \mathcal{P}}_\Omega }({\mathbf{X}}) = {{ \mathcal{P}}_\Omega }({\mathbf{M}})~~~~~~
\end{split}
\end{equation}
where $\bar P$ is a penalty for low-rank promotion as introduced in section 2.3,
${{\mathcal{P}}_\Omega }:{\mathbb{R}^{m \times n}} \to {\mathbb{R}^{m \times n}}$ denotes projection onto $\Omega$.
In the case of $\bar P({\mathbf{X}}) = {\left\| {\mathbf{X}} \right\|_0} = {\mathrm{rank}}({\mathbf{X}})$,
(33) is a nonconvex rank minimization problem.
A popular convex relaxation method is to approximate the rank function using the nuclear norm,
i.e., $\bar P({\mathbf{X}}) = {\left\| {\mathbf{X}} \right\|_*}$.
It has been shown in [169], [171] that under certain conditions,
a matrix ${\mathbf{M}} \in {\mathbb{R}^{m \times n}}$ of rank $r \le \min \{m,n\}$
can be exactly recovered from a small of its entries by using the nuclear norm.

In the noisy case, $\{ {{\mathbf{M}}_{i,j}={\mathbf{X}}_{i,j}+\epsilon_{i,j}}:(i,j) \in \Omega \} $
where $\epsilon_{i,j}$ is i.i.d. Gaussian noise,
a robust variant of (33) is more favorable as
\begin{equation}
\begin{split}
&\mathop {{\mathrm{min}}}\limits_{\mathbf{X}} \bar P({\mathbf{X}})\\
\mathrm{subject~ to~}&~{\left\| {{{ \mathcal{P}}_\Omega }({\mathbf{X}}) - {{ \mathcal{P}}_\Omega }({\mathbf{M}})} \right\|_\mathrm{F}} \le \sigma
\end{split}
\end{equation}
where $\sigma > 0$ is the noise tolerance.
This constrained formulation (34) can be converted into an unconstrained formulation as
\begin{equation}   
\mathop {{\mathrm{min}}}\limits_{\mathbf{X}} \frac{1}{2}\left\| {{{ \mathcal{P}}_\Omega }({\mathbf{X}}) - {{ \mathcal{P}}_\Omega }({\mathbf{M}})} \right\|_\mathrm{F}^2 + {\bar P_\lambda }({\mathbf{X}})
\end{equation}
where $\lambda> 0$ is the a regularization parameter.

The superiority of a nonconvex regularization (e.g., the Schatten-$q$ norm)
over the nuclear norm has been widely demonstrated in [6], [48], [175]--[179].
In [6], [175], a proximal descent (PD) algorithm has been proposed,
which can be viewed as a special case of the PD algorithm (37) with
${\bar P_\lambda }$ be the Schatten-$q$ norm. In [48],
an iteratively reweighted algorithm for the unconstrained formulation (35)
has been designed via smoothing the Schatten-$q$ norm, which involves solving a sequence of linear equations.
It has been shown in [177] that, for reliable recovery of low-rank matrices from compressed measurements,
the sufficient condition of Schatten-$q$ norm regularization is weaker than that of nuclear norm regularization.
Moreover, robust matrix completion using Schatten-$q$ regularization has been considered in [179].
Meanwhile, in [180], a truncated nuclear norm has been designed to gain performance improvement over the nuclear norm.
More recently, the MCP penalty has been used and an ADMM algorithm has been developed in [181],
whilst an approximated $\ell_0$ penalty has been considered in [182].

Among the above nonconvex methods, subsequence convergence is proved for the methods [6], [48], [175], [176], [181].
In fact, based on the recent convergence results for nonconvex and nonsmooth optimization [69]--[71],
global convergence of the methods [6], [175], [181] can be guaranteed under some mild conditions.

Following the recent results in [69]--[71], efficient and globally convergent
first-order algorithms for the unconstrained problem (35) can be developed for
a generalized nonconvex ${\bar P_\lambda}$. Using PD method for example, let
$F({\mathbf{X}}) = 1/2\left\| {{{ \mathcal{P}}_\Omega }({\mathbf{X}}) - {{ \mathcal{P}}_\Omega }({\mathbf{M}})} \right\|_\mathrm{F}^2$,
consider the following approximation of the objective in (35) at iteration $k+1$ and at a given point ${{\mathbf{X}}^k}$ as
\begin{equation}   
\begin{split}
{Q_{{L_F}}}({\mathbf{X}};{{\mathbf{X}}^k}) = &F({{\mathbf{X}}^k}) + \left\langle {{\mathbf{X}} - {{\mathbf{X}}^k},\nabla F({{\mathbf{X}}^k})} \right\rangle\\
&~~~~ + \frac{{{L_F}}}{2}\left\| {{\mathbf{X}} - {{\mathbf{X}}^k}} \right\|_\mathrm{F}^2 + {\bar P_\lambda }({\mathbf{X}})
\end{split}
\end{equation}
where $\nabla F({{\mathbf{X}}^k}) = {{ \mathcal{P}}_\Omega }({\mathbf{X}}^k) - {{ \mathcal{P}}_\Omega }({\mathbf{M}})$
and ${L_F} > 0$ is a proximal parameter. Then, with the definition of ${\bar P_\lambda }$ in (6),
minimizing ${Q_{{L_F}}}({\mathbf{X}};{{\mathbf{X}}^k})$ reduces to the generalized proximity operator in Theorem 2 (section 2.3) as
\begin{equation}  
{{\mathbf{X}}^{k + 1}} = {\mathrm{pro}}{{\mathrm{x}}_{(1/{L_F}){{\bar P}_\lambda }}}\left( {{{\mathbf{X}}^k} - \frac{1}{{{L_F}}}\nabla F({{\mathbf{X}}^k})} \right)
\end{equation}
which can be computed as (8).

This iterative singular-value thresholding algorithm also fits into the
framework of majorization-minimization (MM) method. From the results in [69]--[71],
a sufficient condition of convergence for this algorithm is given as follows.
Suppose that ${\bar P_\lambda}$ is a closed, proper, lower semi-continuous, KL function,
if ${L_F} > 1$, then, the sequence $\{ {{\mathbf{X}}^k}\}$ generated by (37)
converges to a stationary point of the problem (35). Meanwhile, the efficient ADMM algorithm
also can be applied to the problem (35) with convergence guarantee under some mild conditions [71].

For large-scale matrix completion problems,
matrix factorization is a popular approach
since a matrix factorization based optimization formulation, even with huge size,
can be solved very efficiently by standard optimization algorithms [183].
Very recently, scalable equivalent formulations of Schatten-$q$
quasi-norm have been proposed in [184]--[186],
which facilitate the design of highly efficient algorithms
that only need to update two much smaller factor matrices.

\textit{Example 7 (Low-rank image recovery)}.
We evaluate the algorithm (37) with hard-, $\ell_q$- and soft-thresholding
on gray-scale image recovery on the ``Boat'' image of size $512\times512$,
where 50\% of pixels are observed in the presence of Gaussian noise with SNR = 40 dB, as shown in Fig. 12.
Two cases are considered.
\textit{1) Non-strictly low-rank}: the original image is used,
which is not strictly low-rank with the singular values approximately following an exponential decay,
as shown in Fig. 13.
\textit{2) Strictly low-rank}: the singular values of the original image are truncated and
only the 50 largest values are retained, which results in a new image which is strictly low-rank used for performance evaluation.
The regularization parameter $\lambda$ is selected
by providing the best performance. In implementing the algorithm with hard- and $\ell_q$-thresholding,
we first run it with soft-thresholding to obtain an initialization.

Fig. 12 shows the recovered images along with the relative error of recovery (RelErr).
Fig. 14 shows the performance of the $\ell_q$-thresholding versus $q$ in the two cases.
For the two cases, the best performance of $\ell_q$-thresholding are given by $q=0.8$ and $q=0.2$, respectively.
The results indicate that a relatively small value of $q$ should be used in the strictly low-rank case,
while a relatively large value of $q$ should be used in the non-strictly low-rank case.
The $\ell_0$ and $\ell_q$ penalties can achieve significant performance improvement
over the $\ell_1$ one only in the strictly low-rank case.

\begin{figure*}[!t]
\centering
\subfloat[The original image is used (non-strictly low-rank)]{
{\includegraphics[scale = 0.4]{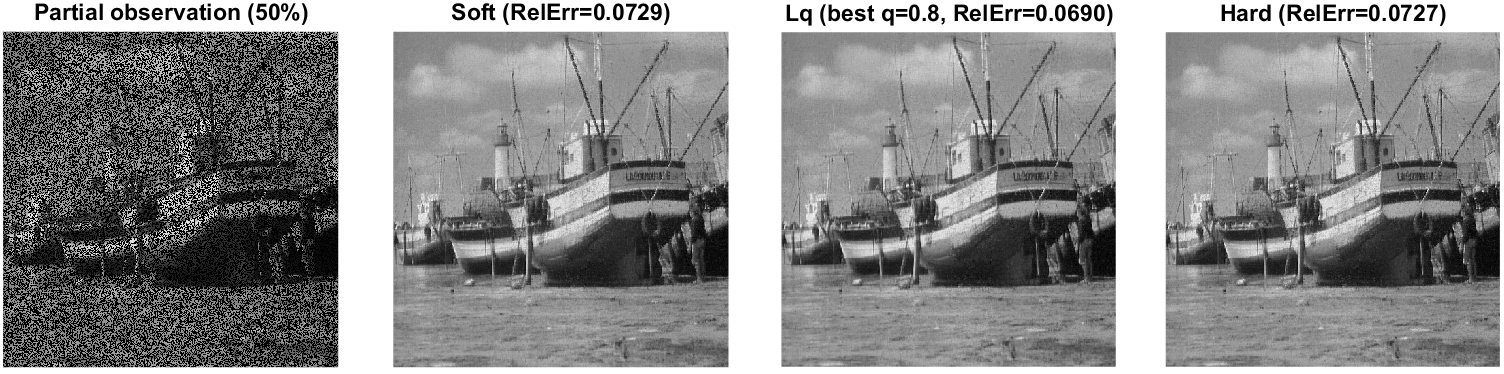}}}\\
\subfloat[The singular values of the original image are truncated and only the 50 largest values are retained (strictly low-rank)]{
{\includegraphics[scale = 0.4]{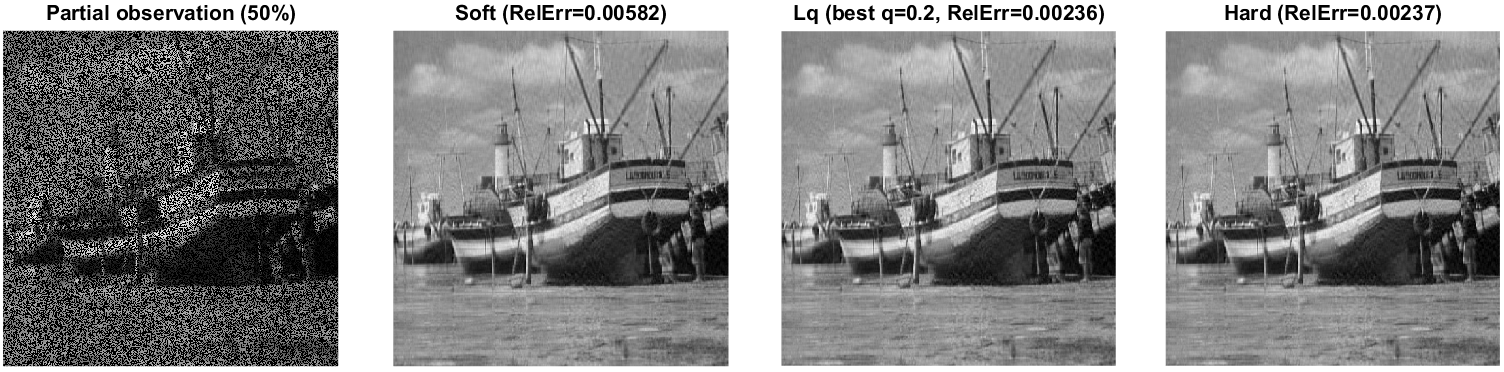}}}
\caption{Recovery performance of the hard-, $\ell_q$- and soft-thresholding in reconstructing an $512 \times 512$ image.}
\label{figure12}
\end{figure*}

\begin{figure}[!t]
 \centering
 \includegraphics[scale = 0.6]{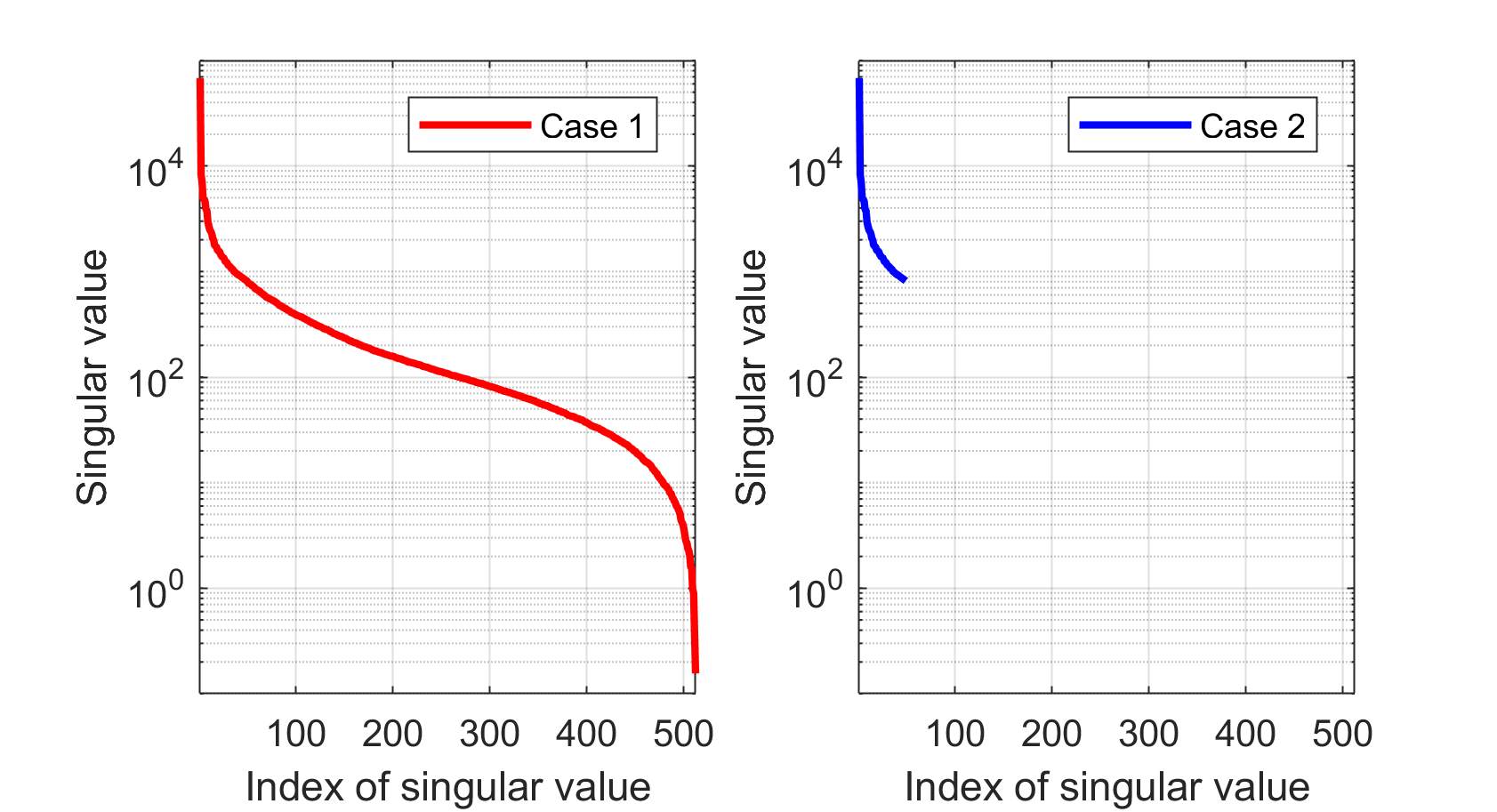}
\caption{Sorted singular values in the two cases.}
 \label{figure13}
\end{figure}

\begin{figure}[!t]
 \centering
 \includegraphics[scale = 0.64]{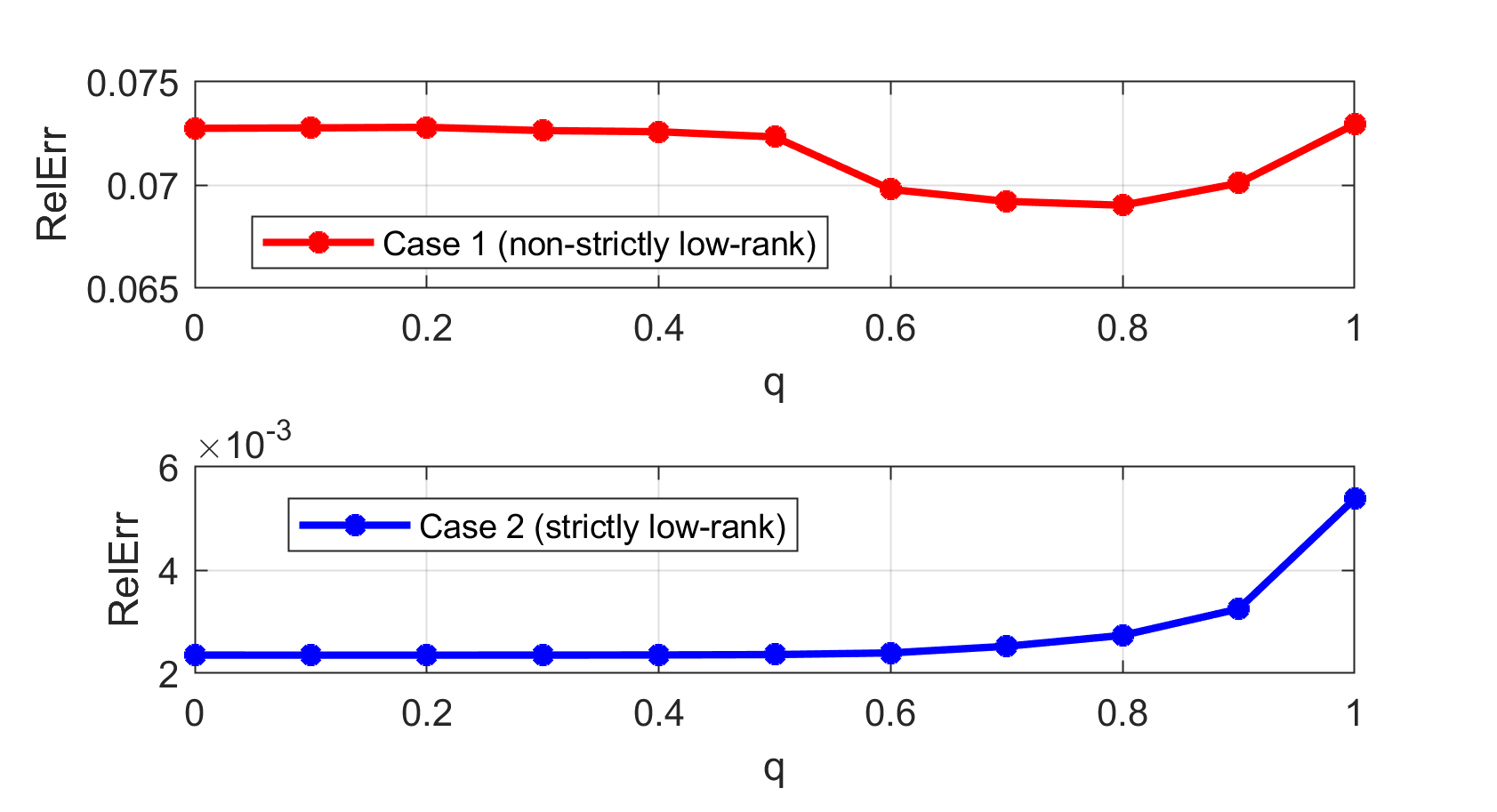}
\caption{Recovery performance of $\ell_q$-thresholding versus $q$.}
 \label{figure14}
\end{figure}

\subsection{Robust PCA}

The objective of robust PCA is to enhance the robustness of PCA against
outliers or corrupted observations [187]--[189]. As these traditional methods
cannot yield a solution in polynomial-time with performance guarantees under mild conditions,
an improved version of robust PCA has been proposed in [190], [191]. In this new version,
the robust PCA problem is treated as a low-rank matrix recovery problem in the presence of sparse corruption,
which in fact is a joint sparse and low-rank recovery problem. Specifically,
the goal is to recover a low-rank matrix ${\mathbf{L}} \in {\mathbb{R}^{m \times n}}$ from highly (sparsely) corrupted observation
\begin{equation}
{\mathbf{M}} = {\mathbf{L}} + {\mathbf{S}}   \notag
\end{equation}
where ${\mathbf{S}} \in {\mathbb{R}^{m \times n}}$ represents the sparse corruption,
in which the entries can have arbitrarily large magnitude and their support can be assumed to be sparse but unknown.
There exist many important applications involving such a low-rank and sparse decomposition problem,
such as video surveillance, face recognition, latent semantic indexing and ranking
and collaborative filtering [191], to name a few.
This can be achieved via exploiting the low-rankness of $\mathbf{M}$ by the following formulation
\begin{equation}
\begin{split}
& \mathop {{\mathrm{min}}}\limits_{{\mathbf{L}},{\mathbf{S}}} {G_1}({\mathbf{L}}) + \lambda {G_2}({\mathbf{S}})~~~~~~~~~~~~~~~~~~\\
\mathrm{subject~ to~}&~ {\mathbf{M}} = {\mathbf{L}} + {\mathbf{S}}
\end{split}
\end{equation}
where ${G_1}$ and ${G_2}$ are penalties for low-rank and sparsity promotion, respectively.
With ${G_1}({\mathbf{L}}) = {\left\| {\mathbf{L}} \right\|_*}$ and ${G_2}({\mathbf{S}}) = {\left\| {\mathbf{S}} \right\|_1}$,
(38) becomes the principal component pursuit (PCP) formulation [191].
It has been shown in [191] that under rather weak conditions,
exactly recovery of the low-rank ${\mathbf{L}}$ and the sparse ${\mathbf{S}}$ can be achieved by the convex PCP formulation.

Consider both small entry-wise noise and gross sparse errors,
which is a more practical case in many realistic applications,
the formulation (38) can be extended to
\begin{equation}
\begin{split}
& \mathop {{\mathrm{min}}}\limits_{{\mathbf{L}},{\mathbf{S}}} {G_1}({\mathbf{L}}) + \lambda {G_2}({\mathbf{S}})~~~~~~~~~~~~~~\\
\mathrm{subject~ to~}&~ {\left\| {{\mathbf{M}} - {\mathbf{L}} - {\mathbf{S}}} \right\|_\mathrm{F}} \le \sigma
\end{split}
\end{equation}
where $\sigma > 0$ is the noise tolerance.
A more tractable alternative formulation of (39) is the following unconstrained formulation
\begin{equation}   
\mathop {{\mathrm{min}}}\limits_{{\mathbf{L}},{\mathbf{S}}} {G_1}({\mathbf{L}}) + \lambda {G_2}({\mathbf{S}}) + \frac{1}{{2\mu }}\left\| {{\mathbf{M}} - {\mathbf{L}} - {\mathbf{S}}} \right\|_\mathrm{F}^2
\end{equation}
where $\mu > 0$ is the a penalty parameter.
With ${G_1}({\mathbf{L}}) = {\left\| {\mathbf{L}} \right\|_*}$ and ${G_2}({\mathbf{S}}) = {\left\| {\mathbf{S}} \right\|_1}$,
(39) reduces to the stable principal component pursuit in [192], and its variant (40) is solved by accelerated proximal gradient algorithm [192].

Recently, to attain performance improvement over using the convex nuclear norm and $\ell_1$ penalties,
nonconvex penalties have been considered in [9], [193]--[196]. Specifically, in [9], the $q$-shrinkage
regularization has been used for both $\mathbf{L}$ and $\mathbf{S}$ in (40) and an ADMM algorithm has been proposed.
In [193], a nonconvex method for (38) has been developed via alternating projection of the residuals onto
the set of low-rank matrices and the set of sparse matrices. In [194], the capped norm penalty has been
considered in the formulation (39) and an ADMM algorithm has been proposed.
More recently, based on the formulation (40) and using the $\ell_0$ penalty for $\mathbf{S}$ and a low-rank factorization
for $\mathbf{L}$, an  cyclic descent algorithm has been designed in [195]. Meanwhile, using the formulation (38),
an ADMM algorithm employing a rank approximation penalty has been proposed in [196].

Among all these methods, except for the convex methods [191], [192],
there is no global convergence guarantee for the nonconvex algorithms. Following the recent results
in nonconvex and nonsmooth optimization [69], [70], we can develop an efficient and
globally convergent algorithm for generalized nonconvex penalties ${G_1}$ and ${G_2}$, e.g.,
for the formulation (40), based on the block coordinate descent (BCD)
method (also known as alternating minimization) as
\begin{equation}   
\begin{split}
{{\mathbf{L}}^{k + 1}} = &\arg \mathop {\min }\limits_{\mathbf{L}} {G_1}({\mathbf{L}}) \\
& + \frac{1}{{2\mu }}\left\| {{\mathbf{M}} - {\mathbf{L}} - {{\mathbf{S}}^k}} \right\|_\mathrm{F}^2 + \frac{{{c_k}}}{2}\left\| {{\mathbf{L}} - {{\mathbf{L}}^k}} \right\|_\mathrm{F}^2
\end{split}
\end{equation}
\begin{equation}   
\begin{split}
{{\mathbf{S}}^{k + 1}} = &\arg \mathop {\min }\limits_{\mathbf{S}} \lambda {G_2}({\mathbf{S}}) \\
& + \frac{1}{{2\mu }}\left\| {{\mathbf{M}} - {{\mathbf{L}}^{k + 1}} - {\mathbf{S}}} \right\|_\mathrm{F}^2 + \frac{{{d_k}}}{2}\left\| {{\mathbf{S}} - {{\mathbf{S}}^k}} \right\|_\mathrm{F}^2
\end{split}
\end{equation}
where ${c_k} > 0$ and ${d_k} > 0$.
These two subproblems can be efficiently solved by the proximity operator introduced in section 2.

This BCD algorithm considers the proximal regularization of the Gauss-Seidel scheme
by coupling the Gauss-Seidel iteration scheme with a proximal term.
Using this proximal regularization strategy, it can be derived following the results in [69], [70] that,
when ${G_1}$ and ${G_2}$ are closed, proper, lower semi-continuous and KL functions,
the sequence $\{ {{\mathbf{L}}^k},{{\mathbf{S}}^k}\}$ generated via (41) and (42) converges to a stationary point of the problem (40).
Meanwhile, the ADMM method can also be applied to solve (40)
with guaranteed convergence under some mild conditions [197].

\section{Other Applications Involving Nonconvex Sparse and Low-Rank Regularization}

The goal of this article is to provide a comprehensive overview
on nonconvex regularization based sparse and low-rank recovery.
In a field as wide as this, except for the applications introduced
above, in this section we further briefly review some other applications
where nonconvex regularized sparse and low-rank recovery has been applied.

\textit{Subspace Learning and Tracking}:
In addition to the matrix completion and robust PCA problems introduced above,
nonconvex regularization based low-rank recovery has also been used in
subspace learning and tracking, which has applications in
pattern recognition (e.g., object tracking, activity recognition and video surveillance)
and image classification. For such applications, the Schatten-$q$ norm has been used to
achieved better subspace learning and tracking in [198], [199].

\textit{Dictionary Learning for Sparse Coding}:
Dictionary learning for sparse coding aims to
learn an overcomplete dictionary on which the input data can be succinctly represented,
i.e., a linear combination of only a few atoms of the learned dictionary.
It has wide applications in signal/image processing, computer vision, and machine learning [228].
The $\ell_0$ norm penalty has been widely used in dictionary learning for sparse coding [229].
For general non-convex spare coding problems,
alternating algorithms with established convergence guarantee
have been proposed in [214].
More recently, the $\ell_{1/2}$ penalty and the log penalty have
been employed for dictionary learning in [213], [215].

\textit{Nonconvex Regularizers with Redistributing Nonconvexity}:
To facilitate the efficient solving of nonconvex regularized problems,
a nonconvexity redistributing method has been proposed recently in [216].
The core idea is to move the  nonconvexity associated with a nonconvex
penalty to the loss term. The nonconvex regularization
term is convexified to be a convex one, e.g., the $\ell_1$ norm, whilst
the augmented loss term maintains the Lipschitz smooth property.
In such a manner, the transformed problem can be efficiently solved by
well-developed existing algorithms designed for convex regularized problems.

\textit{Maximum Consensus Robust Fitting in Computer Vision}:
Robust model fitting is a fundamental problem in many computer
vision applications, where it is needed to deal with real-life raw data,
e.g., in multi-view geometry and vision-based
localization in robotics navigation. For robust
model fitting, the maximum consensus criterion is of the most
popular and useful, which is in fact a linearly
constrained $\ell_0$ minimization problem.
Deterministic algorithms for the maximum consensus $\ell_0$ minimization
have been proposed recently in [217], [230], which showed superior performance
in both solution quality and efficiency.

\textit{Image Deconvolution and Restoration}:
For the image deconvolution and restoration application,
${\ell _q}$ regularization has been used in [121]--[123]
to attain improved performance over the ${\ell _1}$ regularization.
Meanwhile, ${\ell _0}$ norm combined with total variation has been
considered in [119] to achieve robust restoration in the presence of
impulsive noise.

\textit{Least-Mean-Square (LMS) Filter}:
For sparse system identification, regularized least-mean-square (LMS) algorithms
have shown advantage over traditional LMS algorithms,
e.g., be more accurate, more efficient and more robust to additive noise.
$\ell_0$ constrained LMS algorithms have been designed and analyzed in [224], [225].
Then, a weighted $\ell_2$ regularization based
normalized LMS filter has been proposed in [219],
with application to acoustic system identification and active noise control.
Moreover, $\ell_p$ regularization with $0<p\leq1$ has been considered in [226], [227].

\textit{Simultaneously Sparse and Low-Rank Matrix Recovery}:
While robust PCA introduced in the last section aims to
decompose a low-rank component form sparsely corrupted observation,
the works [220]--[223], [231] consider the recovery of matrices which are
simultaneously sparse and low-rank.
For simultaneously sparse and low-rank recovery, it has been shown in
[222] that nonconvex formulations can achieve reliable performance with
less measurements than convex formulations.
In [220], an ADMM algorithm using an iteratively reweighted $\ell_1$ scheme has been
proposed and applied to hyperspectral image unmixing.
More recently, a nonconvex and nonseparable regularization method
derived based on the sparse Bayesian learning framework
has been presented in [223].

\textit{Matrix Factorization Based Low-Rank Recovery}:
In addition to the general low-rank recovery models
introduced in the last section, there also exist
a class of low-rank models based on low-rank matrix factorization [203], [204].
A significant feature (advantage) of such matrix factorization based
methods is that, matrix factorization enables the algorithms to
scale well to large-scale problems.
Although matrix factorization makes the related formulation
nonconvex (more precisely biconvex), it has been proven that
such formulation for matrix completion has no spurious local minima,
i.e., all local minima are also global [183], [232].

\section{Conclusion and discussion}

In this overview paper, we have presented recent developments of nonconvex regularization
based sparse and low-rank recovery in various fields in signal/image processing, statistics
and machine learning, and addressed the issues of penalty selection, applications and the
convergence of nonconvex algorithms. In recent, nonconvex regularization has attracted much
study interest and promoted the progress in nonconvex and nonsmooth optimization. As a result,
for many applications, convergent and efficient first-order algorithms have been developed for
nonconvex regularized problems.

As shown in many applications, a nonconvex penalty can achieve significant performance improvement
over the $\ell_1$ norm penalty. However, there exist certain instances where the use of nonconvex
regularization will not significantly improve performance, e.g., when the signal is not strictly sparse
(or the matrix is not strictly low-rank) and/or the SNR is low. In such a case, the use of nonconvex
regularization may be unnecessary, considering that the related nonconvex optimization problems
are less tractable than convex problems. Specifically, for a nonconvex regularized algorithm,
the performance is closely related to the initialization and the convergence rate is usually slower
than that of a convex regularized algorithm.

Although it is difficult to determine the best selection of the penalty for a special instance,
it can be selected in an application dependent manner.
Specifically, from the results in the experimental examples,
when the intrinsic component is strictly sparse (or has relatively high sparsity) and the noise is relatively low,
a penalty with aggressive thresholding function (e.g., the $\ell_q$ norm with a relatively small value of $q$)
should be used. Whereas, when the intrinsic component is non-strictly sparse (or has relatively low sparsity)
and/or the noise is relatively high, a penalty with less aggressive thresholding function
(e.g., the $\ell_q$ norm with a relatively large value of $q$) tend to yield better performance.
The same philosophy applies to the low-rank recovery problems,
depends on whether the intrinsic component is strictly low-rank or not.

For the nonconvex and nonsmooth problems reviewed in this paper,
first-order algorithms are usually of the most efficient,
such as the proximal gradient descent, block coordinate descent, and ADMM algorithms.
The dominant computational complexity of such algorithms in each iteration is matrix-vector
multiplication for the sparse recovery problems, e.g., the problems in section III.
Meanwhile, the dominant computational complexity of these first-order algorithms
in each iteration is SVD calculation for the low-rank recovery problems, e.g.,
the matrix completion and robust PCA problems in section V. However,
the theoretical convergence rate of a nonconvex algorithm is generally difficult to derive.
To the best of our knowledge, local (eventually) linear convergence rate has been established
only for the proximal gradient descent algorithm for some special penalties
with discontinuous threshoding functions, such as the hard and $\ell_q$ thresholding [73].
For other algorithms with a general nonconvex penalty,
the theoretical convergence rate is still an open problem.

\appendices

\section{Proof of Theorem 1}
Denote
\begin{equation}
f({\mathbf{x}}) = {P_\lambda }({\left\| {\mathbf{x}} \right\|_2}) + \frac{1}{2}\left\| {{\mathbf{x}} - {\mathbf{t}}} \right\|_2^2. \notag
\end{equation}
By simple geometrical arguments, we first show that a minimizer ${{\mathbf{x}}^ * }$ of $f$
satisfies that ${{\mathbf{x}}^ * } = \alpha {\mathbf{t}}$ for some $\alpha \ge 0$. Specifically,
assume that ${\left\| {{{\mathbf{x}}^ * } - {\mathbf{t}}} \right\|_2} = r$ and consider the set
$\Omega  = \{ {\mathbf{x}}:{\left\| {{\mathbf{x}} - {\mathbf{t}}} \right\|_2} = r\} $,
the points in the set $\Omega$ are lying on the ball with center at ${\mathbf{t}}$ and radius $r$.
Since ${P_\lambda}$ is a non-decreasing function in $[0,\infty )$, in the set $\Omega$,
a minimal value of ${P_\lambda }({\left\| \cdot \right\|_2})$ is given by the point
which is the intersection of the ball and the vector $\mathbf{t}$. Thus, ${{\mathbf{x}}^*} = \alpha {\mathbf{t}}$ with some $\alpha \ge 0$.
Using this property, we have
\begin{equation}
f({{\mathbf{x}}^*}) = {P_\lambda }({\left\| {\mathbf{t}} \right\|_2}\alpha ) + \frac{1}{2}\left\| {\mathbf{t}} \right\|_2^2{(\alpha  - 1)^2}.  \notag
\end{equation}
Further, $\alpha$ should be the minimizer of the function
\begin{equation}
h(\alpha ) = {P_\lambda }({\left\| {\mathbf{t}} \right\|_2}\alpha ) + \frac{1}{2}\left\| {\mathbf{t}} \right\|_2^2{(\alpha- 1)^2},\notag
\end{equation}
which is given by 0 when ${\left\| {\mathbf{t}} \right\|_2} = 0$ and
otherwise given by ${\mathrm{pro}}{{\mathrm{x}}_{{P_\lambda }}}({\left\| {\mathbf{t}} \right\|_2})/{\left\| {\mathbf{t}} \right\|_2}$, which results in (5).

It is easy to see that, when $\mathbf{t}$ is a scalar, i.e., $L=1$, (5) reduces to the
proximity operator (1) for a scalar. For the $\ell_1$ penalty, i.e.,
${P_\lambda }(\cdot) = \lambda {\left\|\cdot  \right\|_1}$, we have
\begin{equation}  
{\mathbf{pro}}{{\mathbf{x}}_{{P_\lambda }}}({\mathbf{t}}) = {\mathbf{t}} \cdot \max \left( {1 - \frac{\lambda}{{{\left\| {\mathbf{t}} \right\|}_2}},0} \right).
\end{equation}
Meanwhile, for the ${\ell_0}$ penalty, i.e., ${P_\lambda }(\cdot) = \lambda {\left\| \cdot \right\|_0}$,
the proximity operator becomes
\begin{equation}  
{\mathbf{pro}}{{\mathbf{x}}_{{P_\lambda }}}({\mathbf{t}}) = \left\{ {\begin{array}{*{20}{l}}
{{\mathbf{0}},}&{{{\left\| {\mathbf{t}} \right\|}_2} < \sqrt {2\lambda } }\\
{\{ {\mathbf{0}},{\mathbf{t}}\} ,}&{{{\left\| {\mathbf{t}} \right\|}_2} = \sqrt {2\lambda } }\\
{{\mathbf{t}},}&{{{\left\| {\mathbf{t}} \right\|}_2} > \sqrt {2\lambda } }
\end{array}} \right. .
\end{equation}

\section{Proof of Theorem 2}
Without loss of generality, we assume that $m \ge n$.
Let ${\mathbf{X}} = {\mathbf{SD}}{{\mathbf{E}}^T}$ be an SVD of ${\mathbf{X}}$,
where ${\mathbf{D}} = {\mathrm{diag}}({d_1}, \cdots ,{d_n}) \in {\mathbb{R}^{n \times n}}$
contains the singular values, ${\mathbf{S}} = [{{\mathbf{s}}_1}, \cdots ,{{\mathbf{s}}_n}] \in {\mathbb{R}^{m \times n}}$
and ${\mathbf{E}} = [{{\mathbf{e}}_1}, \cdots ,{{\mathbf{e}}_n}] \in {\mathbb{R}^{n \times n}}$
contain the left and right singular vectors, respectively.
The objective function in (7) can be expressed as
\begin{align}
&{{\bar P}_\lambda }({\mathbf{X}}) + \frac{1}{2}\left\| {{\mathbf{X}} - {\mathbf{M}}} \right\|_{\mathrm{F}}^2  \notag \\
& = \sum\limits_{i = 1}^n {{P_\lambda }({d_i})}  + \frac{1}{2}\left( {\sum\limits_{i = 1}^n {d_i^2}  - 2\sum\limits_{i = 1}^n {{d_i}{\mathbf{s}}_i^T{\mathbf{M}}{{\mathbf{e}}_i}}  + \left\| {\mathbf{M}} \right\|_{\mathrm{F}}^2} \right).    \notag
\end{align}
Then, the formulation (7) can be equivalently expressed as
\begin{align} 
{\mathrm{pro}}{{\mathrm{x}}_{{{\bar P}_\lambda }}}({\mathbf{M}}) = & \arg \mathop {\min }\limits_{{\mathbf{S}},{\mathbf{D}},{\mathbf{E}}} \sum\limits_{i = 1}^n {\frac{1}{2}d_i^2 - {d_i}{\mathbf{s}}_i^T{\mathbf{M}}{{\mathbf{e}}_i} + {P_\lambda }({d_i})}   \notag  \\
&\mathrm{subject~ to~~~}  {{\mathbf{S}}^T}{\mathbf{S}} = {{\mathbf{I}}_n}, {{\mathbf{E}}^T}{\mathbf{E}} = {{\mathbf{I}}_n} \notag \\
&~~~~~~~~\mathrm{ and} ~~~ {d_i} \ge 0~~   \mathrm{for} ~~ i = 1, \cdots ,n.
\end{align}
Since the $i$-th summand in the objective in (45) is only dependent on the ${{\mathbf{s}}_i}$,
${{\mathbf{e}}_i}$ and ${d_i}$, the minimization in (45) can be equivalently reformulated as
\begin{align} 
& \mathop {\min }\limits_{{{\mathbf{s}}_i},{{\mathbf{e}}_i},{d_i} \ge 0} \frac{1}{2}d_i^2 - {d_i}{\mathbf{s}}_i^T{\mathbf{M}}{{\mathbf{e}}_i} + {P_\lambda }({d_i}) \notag\\
\mathrm{subject~ to}~ ~&{\mathbf{s}}_i^T{{\mathbf{s}}_j} = {\delta _{i,j}},~ {\mathbf{e}}_i^T{{\mathbf{e}}_j} = {\delta _{i,j}},~ i,j \in \{ 1, \cdots ,m\}.
\end{align}
For fixed ${d_i} \ge 0$, minimizing the objective in (46) is equivalent to
maximizing ${\mathbf{s}}_i^T{\mathbf{M}}{{\mathbf{e}}_i}$ with respect to ${{\mathbf{s}}_i}$ and ${{\mathbf{e}}_i}$.

Let ${\tilde{\mathbf{ s}}_1}, \cdots ,{\tilde{\mathbf{ s}}_{i - 1}}$ and
${\tilde{\mathbf{ e}}_1}, \cdots ,{\tilde{\mathbf{ e}}_{i - 1}}$ be any left and right singular
vectors of ${\mathbf{M}}$ respectively corresponding to the singular values
${\sigma _1}({\mathbf{M}}) \ge {\sigma _2}({\mathbf{M}}) \ge  \cdots  \ge {\sigma _{i - 1}}({\mathbf{M}})$,
consider the following fact that, for any $i \ge 1$ the problem
\begin{align}
&\mathop {\max }\limits_{{{\mathbf{s}}_i},{{\mathbf{e}}_i}} {\mathbf{s}}_i^T{\mathbf{M}}{{\mathbf{e}}_i}   \notag \\
\mathrm{subject~ to~~} &{{\mathbf{s}}_i} \bot \{ {\tilde{\mathbf{ s}}_1}, \cdots ,{\tilde{\mathbf{ s}}_{i - 1}}\},~ \left\| {{{\mathbf{s}}_i}} \right\|_2^2 \le 1 \notag\\
\mathrm{and}~~  & {{\mathbf{e}}_i} \bot \{ {\tilde{\mathbf{ e}}_1}, \cdots ,{\tilde{\mathbf{ e}}_{i - 1}}\}, ~\left\| {{{\mathbf{e}}_i}} \right\|_2^2 \le 1 \notag
\end{align}
is solved by ${\tilde{\mathbf{ s}}_i}$ and ${\tilde{\mathbf{ e}}_i}$,
the left and right singular vectors of ${\mathbf{M}}$ corresponding to the
$i$-th largest singular value ${\sigma_i}({\mathbf{M}})$ (this result is well established in PCA).
With such a solution, the formulation (46) becomes
\begin{equation}   
\mathop {\min }\limits_{{d_i} \ge 0} \frac{1}{2}d_i^2 - {\sigma _i}({\mathbf{M}}){d_i} + {P_\lambda }({d_i}).
\end{equation}
This problem is a form of the proximity operator (1), and the solution is given by
$d_i^* = {\mathrm{pro}}{{\mathrm{x}}_{{P_\lambda }}}\left( {{\sigma _i}({\mathbf{M}})} \right)$,
which consequently results in (8).

\end{document}